\documentclass[usenatbib]{mnras}

\usepackage{newtxtext,newtxmath}
\usepackage[T1]{fontenc}
\usepackage{ae,aecompl}

\usepackage{blindtext}
\usepackage[noabbrev]{cleveref}
\usepackage[utf8]{inputenc}
\usepackage[T1]{fontenc}
\usepackage{graphicx}	% Including figure files
\usepackage{subcaption}
\usepackage{amsmath}	% Advanced maths commands
\usepackage{mathrsfs}
\usepackage{amsfonts}
\usepackage{xcolor}
\newcommand{\abs}[1]{\left\lvert#1\right\rvert}
\title[Magnetic field effects on star formation modelling]{How non-thermal pressure impacts the modelling of star formation in galaxy formation simulations}
\author[Batziou et al.]{
Eirini Batziou,$^{1,2}$\thanks{E-mail: batziou@mpa-garching.mpg.de}
Ulrich P. Steinwandel,$^{3}$
Klaus Dolag$^{2,1}$ and Milena Valentini$^{4,5,6,2}$
\\
$^{1}$Max Planck Institute for Astrophysics, Karl-Schwarzschild-Str. 1, 85748, Garching, Germany\\
$^{2}$Universit\"ats-Sternwarte M\"unchen, Fakult\"at f\"ur Physik, LMU Munich, Scheinerstr. 1, 81679, Germany\\
$^{3}$Center for Computational Astrophysics, Flatiron Institute, 162 5th Avenue, New York, NY 10010, USA\\
$^{4}$Astronomy Unit, Department of Physics, University of Trieste, via Tiepolo 11, I-34131 Trieste, Italy\\
$^{5}$INAF - Osservatorio Astronomico di Trieste, via Tiepolo 11, I-34131 Trieste, Italy\\
$^{6}$ICSC - Italian Research Center on High Performance Computing, Big Data and Quantum Computing}

\date{Accepted XXX. Received YYY; in original form ZZZ}

\pubyear{2022}

\begin{document}
\label{firstpage}
\pagerange{\pageref{firstpage}--\pageref{lastpage}}
\maketitle

\begin{abstract}

In cosmological simulations of large-scale structure star formation and feedback in galaxies are modelled by so-called sub-grid models, that represent a physically motivated approximation of processes occurring below the resolution limit. However, when additional physical processes are considered in these simulations, for instance, magnetic fields or cosmic rays, they are often not consistently coupled within the descriptions of the underlying sub-grid star formation models.
Here, we present a careful study on how one of the most commonly used sub-grid models for star formation in current large-scale cosmological simulations can be modified to self consistently include the effects of non-thermal components (e.g., magnetic fields) within the fluid. We demonstrate that our new modelling approach, that includes the magnetic pressure as an additional regulation on star formation, can reproduce global properties of the magnetic field within galaxies in a setup of an isolated Milky Way-like galaxy simulation, but is also successful in reproducing local properties such as the anti-correlation between the local magnetic field strength with the local star formation rate as observed in galaxies (i.e. NGC 1097).  
This reveals how crucial a consistent treatment of different physical processes is within cosmological simulations and gives guidance for future simulations.

\end{abstract}


%%%%%%%%%%%%%%%%%%%%%%%%%%%%%%%%%%%%%%%%%%%%%%%%%%

%%%%%%%%%%%%%%%%% BODY OF PAPER %%%%%%%%%%%%%%%%%%

\section{Introduction}
Numerical simulations of structure formation are essential for understanding the dynamics of physical processes in the Universe. Usually, they describe scales of Giga-parsec (Gpc) and include galaxy clusters, the intracluster medium (ICM), groups of galaxies, galaxies and their constituents. Modelling different physical processes in such a variety of spatial and mass scales is a challenge that brings the necessity to use physically motivated models describing the physics on scales that are not possible to resolve given the current computational resources. 

Understanding galaxy formation and evolution has been a long-standing problem, theoretically approached by semi-analytic methods \citep[e.g.][]{ Lacey1991,Kauffmann1994}, hydrodynamical simulations of galaxy groups \citep[e.g.,][]{hirschmann2014cosmological,Vogelsberger2014,Schaye2015, Pillepich2018,Crain2023groups}, and hydrodynamical simulations of single galaxies \citep[e.g.,][]{2008Governato,wang2015nihao, Grand2017,hopkins2014galaxies,oser2010two,guedes2011forming,aumer2013towards,marinacci2014formation,murante2015simulating,agertz2016impact,valentini2017effect,Valentini2018,Valentini2023,hopkins2015new,Giammaria2021disks,Hopkins2018fire,Hopkins2022,Hopkins2023}.

Magnetic fields are ubiquitous in the Universe, and observations of disk galaxies show the presence of magnetic fields in (rough) energy equipartition with the remaining energy components of the interstellar medium (ISM) \cite[see, e.g.,][for a review]{BeckReview}.
Therefore, there is a growing effort to involve magnetic fields in simulations of galaxies \citep[e.g.,][]{Wang2009mhd, Dubois2010,pakmor2013simulations, Rieder2016,Butsky2017,Ulli, Ulli-Bwinds,Wibking2023BGal} and in cosmological zoom-in simulations \citep[e.g.,][]{Beck2012,Rieder2017a,Rieder2017b,Pakmor2017,Su2017,Martin-Alvarez2018,Hopkins2020zoom,Pakmor2024,Martin-Alvarez2021}.
Meanwhile, single large-scale cosmological simulations have started to investigate the magnetic field structure on larger scales \citep[e.g.,][]{Pillepich2018,Marinacci2018,Ramesh2023MNRAS}. 

Star formation and stellar feedback are the key ingredients that shape the physics of the ISM of a galaxy, initially studied by \citet[][]{mckee1977theory} and followed up by many groups \citep[e.g.][]{Draine2011bookISM,KimOstriker2015,WalchNaab2015,Gatto2015,Martizzi2015,Haid2016,Lucas2020,Bieri2023,Hirashima2023}.
Numerically, incorporating all the relevant physics in one model for star formation and feedback has long been a challenge.

The first simulations that include star formation considered a simple over-density criterion of a single fluid gas, ignoring the multi-phase nature of the ISM \citep[]{WhiteRees1978, CenOstriker1992, NavarroWhite1993, katz1996cosmological}. Despite the simplicity of these models they have been used to understand the formation of galaxies in general \citep[e.g.,][]{KatzGunn1991,NavarroWhite1994,Steinmetz1995,MihosHernquist1996}.
However, with the single fluid approach, these models resulted in the well-known over-cooling problem due to the lack of stellar feedback that resulted in an overproduction of stars \citep[see, e.g.,][for a detailed discussion]{Somerville2015, Naab2017}.

The feedback of massive stars (i.e. stellar explosions, stellar winds, etc.) is a crucial component of the ISM due to the metal enrichment, builds up turbulence in the medium and disrupts cold clouds in the neighbour of a star \citep[see, e.g.,][for reviews]{Elmegreen2004,Scalo2004,Naab2017}. Stellar feedback and radiative cooling processes in the ISM give rise to a multi-phase medium \citep{mckee1977theory} that consists of hot gas, generated by stellar feedback, warm gas at the cooling/heating equilibrium and cold gas in the cooling dominated regime that can be split in neutral and molecular gas, from which the latter is responsible for the formation of stars. This multi-phase nature of the ISM makes it necessary to modify the star formation recipes and incorporate the multi-phase structure of the ISM in the numerical modelling \citep[e.g.,][]{springel2003cosmological}.

However, in addition to the thermal component of the ISM, other factors influence its structure, such as turbulence, cosmic rays, and magnetic fields.
These ingredients are fundamental components of the energy budget of the ISM but are very often ignored, especially when it comes to large-scale structure formation simulations or are not coupled to the sub-grid models for star formation or ISM modelling. 

Recently, a number of different star formation recipes have been developed to capture processes that are relevant to the multi-phase structure of the ISM. These models can efficiently be coupled to the heating from supernovae (SNe) and account for a treatment of non-equilibrium cooling \citep[e.g.,][]{Krumholz2005,Hopkins2012,Hu2017,Ebagezio2023MNRAS}. Furthermore, there is a class of models that incorporate turbulence \citep[e.g.,][]{Federrath2012, Semenov2018, Kretschmer2020} and models that shift the modelling from a density threshold that initiates star formation to a pressure threshold \citep[e.g.,][]{murante2010subresolution}. Although the models mentioned above present a significant advancement in the treatment of the ISM, they still miss more complicated ingredients, such as the impact of magnetic fields and cosmic rays, which can be easily accounted for in pressure-based models, \cite[but see, e.g.,][for an extension of the turbulence models to account for magnetic fields]{Girma-Teyssier2024}. 

Magnetic fields are present on every scale of the Universe and are significant when ionised matter is involved. On sub-galactic scales, magnetic fields are the main reason for cosmic ray acceleration and propagation via magnetised shocks and are relevant for star formation \citep[e.g.,][for a review]{BeckReview}. Recent studies \citep[e.g.,][]{pakmor2013simulations,Ulli-Bwinds} show that they could potentially contribute to the formation of galactic outflows.
Observations of galaxies 
\citep[e.g.,][]{2013BeckbookObs,Heesen2023} indicate the existence of an ordered magnetic field structure with a magnitude of a few $\mathrm{\mu}$Gauss and a turbulent field of similar strength. It has also been confirmed that the magnetic energy density of galaxies is of the same order as the thermal component of the ISM \citep[e.g.,][]{Tabatabaei2008,Basu2013,Manna2023}.  
The ISM is a low-$\beta$ plasma (where $\beta$ is the ratio of the thermal pressure over the magnetic pressure), and therefore, the magnetic fields can be dynamically important for the global evolution of galaxies, and it is fundamental that they are accurately modelled in numerical simulations.

Here, we present an updated version of the widely used ISM model for cosmological simulations by \citet{springel2003cosmological}, taking into account the effect of the magnetic fields and allowing for the effect of other non-thermal components of the ISM, such as cosmic rays. This model is particularly designed for the next generation of cosmological simulations and systems that usually do not resolve the details of the ISM of galaxies.

This work aims to include the non-thermal effects of the ISM into the model for a self-consistent treatment of star formation in the field of galaxy formation and evolution. In particular, our updated model includes the molecular fraction of the multi-phase ISM, which is responsible for the star-forming gas, in addition to the cold and hot gas that was included in \cite{springel2003cosmological}. We use observational-driven relations between the molecular fraction of the gas and the total pressure of the galaxy \citep{blitz2006role}, which allows us to include the effect of the magnetic fields via the magnetic pressure as an additional regulation for star formation.
In future studies, other non-thermal pressure components can be included with our approach. Apart from including magnetic field effects in the star formation process by taking into account the magnetic pressure in the formation of cold clouds (i.e., molecular fraction of the ISM), we incorporate the effect of the stellar feedback on the magnetic field seeding of the galaxy. This model is proposed in \citet{beck2013SNseeding} and assumes that SNe explosions seed the magnetic field of a galaxy with no need for a pre-existing primordial field. We couple our novel star-formation model with the one outlined in \citet{beck2013SNseeding} to accurately capture the mutual influence between star-formation processes and magnetic fields in a self-consistent manner.

The paper is structured as follows. Sec.~2 describes the equations of the updated star formation model and the motivation for the selection of free parameters. Sec.~3 briefly describes the numerical methods and the simulation setup for an isolated Milky Way-type galaxy simulation embedded in a realistic environment to test our updated star formation and feedback implementation. In Sec.~4, we show the results of the Milky Way-type galaxy simulation and compare the properties of the magnetic field-star formation rate correlation to observations. We summarize and give conclusions in Sec.~5.

\section{Multiphase model for star formation and feedback}
\label{sec:sfr42}
We present an updated model for star formation and feedback that addresses the multi-phase structure of the ISM and incorporates magnetic fields to regulate the star formation process. Our model is an extension of the widely-used star formation model by \citet{springel2003cosmological} (SH03 hereafter). In addition, we couple our model to the magnetic seeding model of \citet{beck2013SNseeding}, which describes the magnetic field seeding through SNe explosions, which rate is derived from our updated star formation recipe. Therefore, we self-consistently model the effect of the magnetic fields in the star formation process of the galaxy that shapes the ISM and also the effect of the star formation on the magnetic field evolution in galaxies. In \cref{sec:equations_of_model}, we present the equation of our model; in \cref{sec:pars}, we specify and motivate the selection of the free parameters in our model, and in \cref{sec:beck}, we summarize the work of \citet{beck2013SNseeding} and how it is altered based on our model details. 

\subsection{Equations of the model}
\label{sec:equations_of_model}

The new star formation model presented here is a modification of the widely used SH03 model for star formation and feedback and is implemented in the developer's version of the smoothed particle magneto-hydrodynamics (SPMHD) code \textsc{gadget} \citep{springel2001gadget,springel2005gadget,dolag2009mhd}. In our model, a resolution element representing a segment of the ISM encompasses both hot and cold gas. A proportion of the cold gas undergoes a transition to molecular form, which subsequently gives rise to star formation. A unique aspect of our model is incorporating the molecular fraction, which is regulated by pressure, including thermal and non-thermal components of the ISM. The pressure-regulated molecular fraction of the ISM is absent in SH03. Therefore, we modify their model accordingly and derive the equations that describe the sub-grid model for star formation and feedback in this section. 

Our model includes thermal heating of SNe that evaporates the surrounding cold clouds, forms the hot volume filling phase of the ISM and enriches the ambient ISM with metals. In addition, the hot phase is levelled down due to radiative cooling, quickly restoring the cold phase that eventually transfers into the star-forming molecular phase, often neglected in sub-grid modelling for star formation, which makes the modelling more complex compared to previous approaches. Once the gas reaches a critical density threshold, it will enter the star-forming regime. Even though our model still relies on a density threshold for the gas to undergo star formation, this density threshold is regulated by the total pressure of the galaxy that includes thermal and non-thermal components of the ISM, as it will be shown later in this section. 

\begin{figure}
    \centering
    \includegraphics[width=1.05\columnwidth]{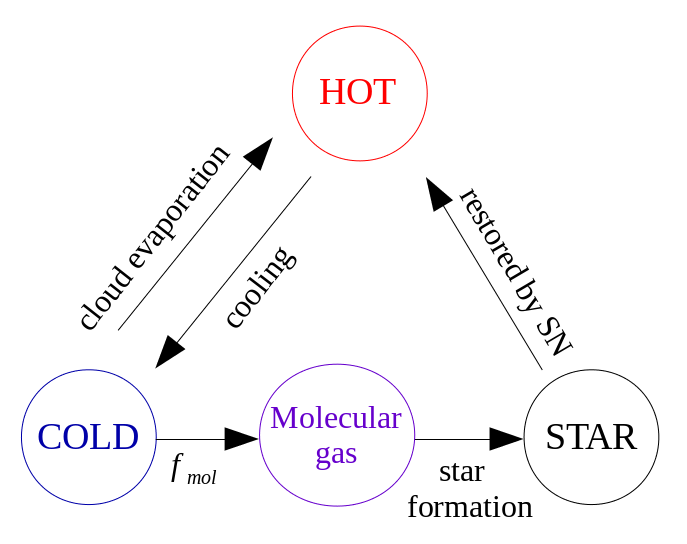}
    \caption{Schematic representation of the mass flow (of a resolution element) between different phases of the ISM and the interaction between them. Hot gas is cooling into the cold phase, from where it transits into the molecular regime before it is turned into stars once it reaches a density threshold. The stars explode as SNe, which restores the hot ISM phase. Furthermore, the hot phase interacts with the cold phase as cold clouds evaporate. The novelty of this model is to include the molecular phase of the ISM, regulated by thermal and non-thermal pressure components of the ISM, which gives rise to star formation.}
    \label{fig:flow}
\end{figure}

A schematic description of the new model can be found in \cref{fig:flow}. The main difference compared to the model of SH03 is altering the multi-phase model in a way that includes an additional component of the ISM, that one of the molecular form. The molecular form of the ISM gas is directly correlated to the total pressure, which involves thermal and non-thermal components of the ISM pressure. We use an observationally driven relation between the pressure and the molecular fraction of a galaxy by \citet{blitz2006role}, which will be further motivated later in this section.

We follow the modelling of SH03 and present here the updated version of the model equations in our three-phase ISM model.
In the updated model, the stars are formed on a characteristic timescale that we choose to be the free fall timescale of the cold gas with a star formation efficiency of $f_*$. Therefore, the star formation rate density is described by the following equation
 \begin{equation}
     \label{eq:1sfr}
     \dfrac{d \rho_*}{dt} = (1-\beta) f_* f_\mathrm{mol} \dfrac{\rho_c}{t_\mathrm{dyn}},
 \end{equation}
with $\rho_*$ being the density of the star-forming gas, $f_\mathrm{mol}$ is the fraction of the gas in molecular form that is star-forming, $\rho_c$ 
\footnote{In all equations of the model, a subscript \textit{c} refers to the quantities describing the cold gas of the medium and \textit{h} refers to the hot gas.} 
is the density of the cold gas, $t_\mathrm{dyn}$ is the dynamical timescale, and $\beta$ is the fraction of the massive, short-lived stars, explode as SN type II in timescales comparable to the dynamical time of the system and, therefore cannot be accounted for the star formation. The parameter $\beta$ depends on the initial mass function (IMF), and here we choose $\beta=0.1$ for a common Salpeter IMF \citep{salpeter1955luminosity}. The star formation rate relates to the gas reservoir of the galaxy through the observational Schmidt-Kennicutt relation \citep{schmidt1959rate, kennicutt1998global}:
\begin{equation}
    \label{eq:schmidt_law}
    \Sigma_\mathrm{SFR} = (2.5 \pm 0.7) \cdot 10^{-4} \left( \dfrac{\Sigma_\mathrm{gas}}{\mathrm{M_\odot pc^{-2}}} \right)^{1.4 \pm 0.15} \mathrm{\dfrac{M_\odot}{yr \cdot kpc^2}}.
\end{equation}
We assume that star formation is associated with the dynamical time of the cold gas, which is the free fall time of the cold clouds given by
\begin{equation}
     \label{eq1t_dyn}
     t_\mathrm{dyn} = \sqrt{\dfrac{3 \pi}{32G\rho_c}}.
\end{equation}
Hereby, $G$ denotes the gravitational constant, and $\rho_\mathrm{c}$ is the (local) density of the cold clouds.

The star formation rate is proportional to the fraction of the molecular clouds $f_\mathrm{mol}$. The molecular fraction of the gas is calculated given the observational relation found in the study of \citet{blitz2006role}. The molecular fraction is proportional to the hydrostatic pressure of the galaxy and is given by
\begin{align}
     \label{eq:1fmol}
     f_\mathrm{mol} = \dfrac{1}{1+P_0/P_\mathrm{ext}},
\end{align}
with $P_0/k_B = 35000\, \mathrm{K cm^{-3}} $ and expresses the external pressure of the ISM when half of the gas is in molecular form and where $k_B$ is the Boltzmann constant. The external pressure $P_\mathrm{ext}$ is calculated from the mid-plane pressure of an infinite disk that consists of gas and stars, assuming that the gas scale height is much lower than the stellar scale height. The pressure is derived from the following expression as shown in \citet{blitz2006role}
\begin{equation}
     \label{eq:press_ext_blitz}
     P_\mathrm{ext} = (2G)^{0.5} \Sigma_g u_g \left[ \rho_\mathrm{stars}^{0.5} + \left(\dfrac{\pi}{4}\rho_g \right) \right].
\end{equation}
In the last equation, $\Sigma_g$ is the surface density of the gas, $u_g$ is the vertical velocity dispersion of the gas, $\rho_\mathrm{stars}$ is the mid-plane density of the stellar population and $\rho_g$ the mid-plane density of the gas. This pressure corresponds to the (total) hydrostatic pressure of the galaxy, which, apart from the thermal pressure, also includes non-thermal components, such as the magnetic field pressure. For the simulations in this work, we approximate this pressure with the SPMHD pressure of the particle plus the magnetic pressure given by the SPMHD code \textsc{gadget}. 

From the stars that are formed, many of them explode as SNe, which heats the surrounding medium in the form of thermal feedback. Thus, the hot gas gains energy at the rate 
\begin{equation}
    \label{eq:1SN_heat}
     \left.\dfrac{d}{dt}(\rho_h u_h) \right\vert_\mathrm{SN} =  \epsilon_\mathrm{SN}  f_*  f_\mathrm{mol}  \dfrac{d \rho_*}{dt} = \beta u_\mathrm{SN} f_* f_\mathrm{mol} \dfrac{d \rho_c}{t_\mathrm{dyn}},
\end{equation}
with $\epsilon_\mathrm{SN}$ the energy per unit mass that corresponds to one SN explosion $\epsilon_\mathrm{SN} = 4 \cdot 10^{48} \, \mathrm{erg \cdot M_\odot ^{-1}}$ for the IMF adopted in this description \citep{salpeter1955luminosity}. In the last equation, $\rho_h$ and $u_h$ refer to the density and energy per unit mass of the hot phase, respectively. 
\begin{figure}
    \centering
    \includegraphics[width=\columnwidth]{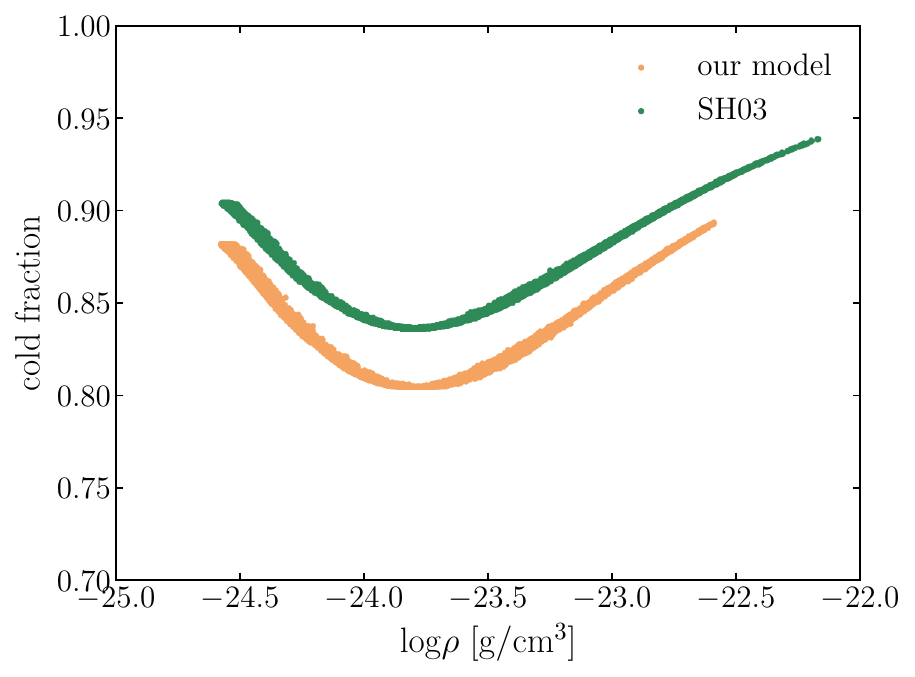}
    \caption{Fraction of the cold phase of the ISM ($x$ in the model equations) as a function of density. The orange points show the result of the new pressure-based star formation model, and the green points represent the outcome of the classic \citet{springel2003cosmological} model for star formation.}
    \label{fig:x_new}
\end{figure}
As a result of the heating, the SN blast wave destroys the star-forming clouds around the region of the SN explosion. Thus, the cold phase evaporated, and the cold mass fraction of the ISM loses mass at the rate
\begin{equation}
    \label{eq:1SN_evap}
    \left.\dfrac{d \rho_c}{dt} \right\vert_\mathrm{EV} =  A \beta f_* f_\mathrm{mol} \dfrac{\rho_c}{t_\mathrm{dyn}},
\end{equation}
with $A$ being the evaporation parameter that scales with density as $A \sim \rho^{-4/5}$ (\citet{mckee1977theory}, see also \cref{sec:pars}). The growth of the cold clouds comes from the radiative cooling of the hot gas. We follow the formulation of SH03 that assumes the thermal instability (TI) operating, giving rise to cold clouds due to the cooling of hot gas. The mass flow from one phase to the other due to radiative cooling is described as 
\begin{equation}
     \label{eq:1thermal_instability}
     \left.\dfrac{d \rho_c}{dt}\right\vert_\mathrm{TI} = - \left.\dfrac{d \rho_h}{dt}\right\vert_\mathrm{TI} = \dfrac{1}{u_h-u_c} \Lambda_\mathrm{net} (\rho_h,u_h),
\end{equation}
with $u_c$ being the energy per unit mass of the cold phase, $u_h$ the energy per unit mass of the hot phase and $\Lambda_\mathrm{net} (\rho_h,u_h) $ the cooling function which is computed from radiative processes taking place in a primordial plasma of Hydrogen and Helium, as presented by \citet{katz1996cosmological}. In \textsc{gadget}, we include collisional excitation of H and He+, ionization of H, He, He+, recombination of H+, He+, He++ and free-free (thermal Bremsstrahlung) emission. The gas cannot cool below $\simeq 10^4 \,\mathrm{K}$ because lower temperatures would require proper treatment of molecular cooling, which can be included in our model but is currently neglected. Thus, we set the temperature of cold clouds to remain constant at $T_c = 10^3\, \mathrm{K}$.

Taking into consideration all the processes being described, the density of the hot and cold phases changes as
\begin{align}
    \label{eq:1mass_rates_cold}
     \dfrac{d \rho_c}{dt} &= - f_* f_\mathrm{mol} \dfrac{ \rho_c}{t_\mathrm{dyn}}-A \beta f_* f_\mathrm{mol} \dfrac{\rho_c}{t_\mathrm{dyn}} + \dfrac{1-f}{u_h-u_c} \Lambda_\mathrm{net}(\rho_h, u_h), \\
      \label{eq:1mass_rates_hot}
      \dfrac{d \rho_h}{dt} &= \beta f_* f_\mathrm{mol} \dfrac{\rho_c}{t_\mathrm{dyn}} + A \beta f_* f_\mathrm{mol} \dfrac{\rho_c}{t_\mathrm{dyn}} - \dfrac{1-f}{u_h-u_c} \Lambda_\mathrm{net}(\rho_h, u_h).
\end{align}
In both differential equations, the first term on the right-hand side represents the gain/loss from the star formation, the second term accounts for the cold cloud evaporation, and the last one is the effect of the thermal instability. The parameter $f$, which can have a value of 1 or 0, represents the onset of the thermal instability. For $f=0$ the thermal instability is operating, and stars are forming. In the opposite case, when $f=1$, ordinary cooling takes place. Following SH03, to physically differentiate the two cases, we use a density threshold, $\rho_\mathrm{thr}$, which will be calculated in \cref{sec:pars}. Thus, star formation takes place for regions where the density of the gas exceeds the density threshold $\rho_\mathrm{thr}$, i.e. when $\rho > \rho_\mathrm{thr}$.

The energy of the total gas changes according to all the processes mentioned above and is described in the equation
\begin{equation}
\begin{split}
    \label{eq:1energy_balance}
    \dfrac{d(\rho_h u_h + \rho_c u_c)}{dt} = -\Lambda_\mathrm{net}(\rho_h, u_h) + &\beta f_* f_\mathrm{mol} \dfrac{\rho_c}{t_\mathrm{dyn}} u_\mathrm{SN} - \\ &(1-\beta)f_* f_\mathrm{mol} \dfrac{\rho_c}{t_\mathrm{dyn}} u_c.
\end{split}
\end{equation}

As mentioned before, the temperature of the cold clouds remains constant at $T_c = 10^3\, \mathrm{K}$. Consequently, we assume that $u_c$ is constant in the above equations, and therefore, we can follow the evolution of the specific energy of the hot phase. 

The hot phase of the gas will evolve according to the differential equation
\begin{equation}
\begin{split}
    \label{eq:1hot_temp_evolution}
    \dfrac{du_h}{dt} &= - \left( \beta \dfrac{f_* f_\mathrm{mol}}{t_\mathrm{dyn}}\dfrac{\rho_c}{\rho_h} + \dfrac{A \beta}{t_\mathrm{dyn}}\dfrac{\rho_c}{\rho_h}  \right) u_h + 
     \\ &f_* f_\mathrm{mol} \dfrac{\beta}{t_\mathrm{dyn}}\dfrac{\rho_c}{\rho_h} (u_\mathrm{SN}+u_c) +f_*f_\mathrm{mol}\dfrac{A \beta}{t_\mathrm{dyn}}\dfrac{\rho_c}{\rho_h}u_c.
\end{split}
\end{equation}
The last equation results in an equilibrium solution for timescales larger than the dynamical time. Thus, the temperature of the hot phase will evolve towards an equilibrium state, as described by
\begin{equation}
    \label{eq:1eq_solution}
    u_h = \dfrac{u_\mathrm{SN}}{A+1} + u_c.
\end{equation}
Deviations from this equilibrium solution are decaying on a timescale 
\begin{equation}
    \label{eq:1timescale_decay}
    \tau_h = \dfrac{t_\mathrm{dyn} \rho_h}{\beta(A+1)f_* f_\mathrm{mol}\rho_c}.
\end{equation}
It is interesting to notice that this timescale depends not only on the cold fraction of the gas but also on the molecular fraction of the cold gas. Therefore, when the magnetic fields are amplified, the molecular fraction rises (as the total pressure rises) and adjusts the equilibrium state compared to the one that does not include magnetic fields.
The growth of cold clouds is balanced by the star formation and SN feedback towards an equilibrium which self-regulates these processes during galaxy evolution. The gas then behaves as an effective medium with pressure
\begin{equation}
    \label{eq:1eff_pressure}
    P_\mathrm{eff} = (\gamma-1)(\rho_h u_h + \rho_c u_c).
\end{equation}
The effective pressure remains constant due to equilibrium. Taking the above into account \cref{eq:1energy_balance} yields
\begin{equation}
    \label{eq:1rho_c_over_t}
    \dfrac{\rho_c}{t_\mathrm{dyn}} = \dfrac{\Lambda_\mathrm{net} (\rho_h,u_h)}{ f_*f_\mathrm{mol}(\beta u_\mathrm{SN} - (1-\beta)u_c)}.
\end{equation}
We can derive the cold fraction of the gas as a function of the gas density. The cold fraction is defined as $x = \dfrac{\rho_c}{\rho}$. The cooling function of the hot phase is $\Lambda_\mathrm{net}(\rho_h,u_h) = (\rho_h/\rho)^2 \Lambda_\mathrm{net}(\rho,u_h)$. Using the definition 
\begin{equation}
    \label{eq:y(x)}
    y(x) = \dfrac{t_\mathrm{dyn} (x) \Lambda_\mathrm{net} (\rho_h,u_h)}{f_*f_\mathrm{mol}(\beta u_\mathrm{SN} - (1-\beta)u_c)},
\end{equation}
the ratio of the densities of the hot and cold phases is written as
\begin{equation}
    \label{eq:1cold_hot_frac}
    \dfrac{\rho_c}{\rho_h} = \dfrac{\rho_h}{\rho} y(x).
\end{equation}

As a result, the cold fraction can be written as a function of the parameter $y$ as
 \begin{equation}
     \label{eq:1x}
     x = 1 +\dfrac{1}{2y(x)}-\sqrt{\dfrac{1}{y(x)}+\dfrac{1}{4y(x)^2}}.
 \end{equation}
In our model, the cold fraction does not depend only on $y$, but on the cold fraction itself since the parameter $y$ is a non-linear function of $x$. The non-linear expression for the cold fraction comes from choosing the dynamical time of the cold gas as a characteristic timescale of the star formation. The dynamical time (see \cref{eq1t_dyn}) depends on the cold fraction $x$ as
\begin{equation}
     t_\mathrm{dyn} = x^{-1/2} \sqrt{\dfrac{3 \pi}{32G \rho}}.
\end{equation}
Given the non-linearity of the dependencies, we can not solve for $x$ but rather find its value numerically. The cold fraction is the root of the function
\begin{equation}
     \label{eq:1function_for_x}
     f(x) = 1 + \dfrac{\sqrt{x}}{2 \delta} - \sqrt{\dfrac{x}{\delta}+\dfrac{x}{2 \delta^2}} - x,
\end{equation}
with $\delta$
\begin{equation}
     \delta = \sqrt{\dfrac{3 \pi}{32 G \rho}} \cdot \dfrac{\Lambda_\mathrm{net} (\rho_h,u_h)}{f_*f_\mathrm{mol} \rho (\beta u_\mathrm{SN} - (1-\beta)u_c)}.
\end{equation}
After testing that the \cref{eq:1function_for_x} has a solution, we numerically solve it with a simple bisection method. The dependence of the cold fraction $x$ on the gas density is shown in \cref{fig:x_new}.
 
For comparison with the star formation model of SH03, we directly compare our updated version to the original model in the same figure. The obvious difference is that, in our case, the cold gas fraction reaches lower values and also does not converge to 1 for high densities. This comes from the dependence of the dynamical time on the cold fraction that inserts a complicated correlation between the cold gas fraction $x$ and the total gas density $\rho$. Thus, for high densities, we don't allow all the gas to be in the cold phase and form more stars, but we have included other regulation factors such as the external pressure of the ISM through the molecular fraction $f_\mathrm{mol}$.
 
In order to distinguish between the different states of the model, i.e. star-forming or non-star-forming, the gas parcel enters the multi-phase model (i.e. star-forming) once its density is higher than a density threshold, $\rho_\mathrm{thr}$ (see discussion on \cref{eq:1mass_rates_cold} and \cref{eq:1mass_rates_hot}). Below this density, we set the cold fraction to $x=0$, which means that the gas is not star-forming. The density threshold is calculated from the model equations and will be discussed in the next section.
 
\subsection{Selection of Parameters}
\label{sec:pars}
\begin{figure}
     \centering
     \includegraphics[width=\columnwidth]{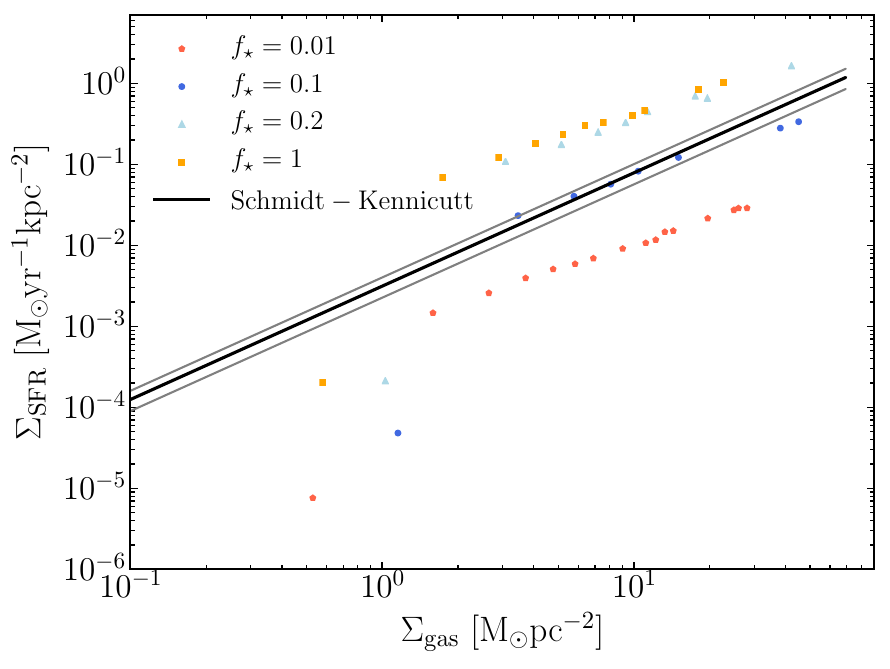}
     \caption{Star formation rate surface density as a function of the surface density of the total gas in a galaxy for different star formation efficiency parameters. The black solid line is the observational Schmidt-Kennicutt relation, and the grey lines represent its errors. All quantities are azimuthically averaged.}
     \label{fig:f_star_param}
\end{figure}
 
The evaporation factor $A$ (introduced in \cref{eq:1SN_evap}) scales with the local density of the ISM, which is theoretically motivated by \cite{mckee1977theory}. We scale it with the density threshold as
\begin{equation}
     \label{eq:1A}
     A(\rho) = A_0 \left( \dfrac{\rho}{\rho_\mathrm{thr}} \right)^{-4/5}.
\end{equation}
In order to constrain $A_0$, we follow SH03 and consider the onset of the thermal instability. From the equilibrium solution in \cref{eq:1eq_solution}, the temperature of the hot phase is approximately $10^5$ K, which is the peak of the cooling function; therefore, the cooling is very effective, and the thermal instability is triggered. Thus $T_\mathrm{SN}/A_0 = 10^5$\,K, which constrains the parameter $A_0$ to be $A_0 = 1000$, for a typical SN temperature of $T_\mathrm{SN}=2 \mu u_\mathrm{SN} m_\mathrm{H} /3k_\mathrm{B} = 10^8$\,K.
 
In the following, we derive the density threshold, which differentiates between star-forming and non-star-forming gas using \cref{eq:y(x)}. 
The cooling function can be written as $\Lambda_\mathrm{net} ( \rho,u_h) / \rho = u_h / t_\mathrm{cool}$. In order to calculate the density threshold, we calculate the value of the cooling function at the baryon over-density calculated as $\rho_\mathrm{ov} = 10^5 \dfrac{3 H^2}{8 \pi G}$, following SH03. Therefore, the density threshold is written as
\begin{equation}
     \label{eq:1densA}
     \rho_\mathrm{thr} = \dfrac{x_\mathrm{thr}}{(1-x_\mathrm{thr})^2} \dfrac{f_\mathrm{mol} (\beta u_\mathrm{SN} - (1-\beta)u_c)}{f_* t_\mathrm{dyn} (\rho_\mathrm{ov}) \Lambda(\rho_\mathrm{ov},u_\mathrm{SN}/A_0)}.
\end{equation}
The cooling function is calculated at the threshold where $u_h = u_\mathrm{SN} /A_0 + u_c \simeq u_\mathrm{SN}/A_0$. The cold fraction at the threshold $x = x_\mathrm{thr}$ is given by $x_\mathrm{thr} = 1 + (A_0+1)(u_c-u_4)/u_\mathrm{SN} \simeq 1 - A_0 u_4/u_\mathrm{SN}$. The latter quantity is calculated by setting the condition $u_\mathrm{eff}(\rho_\mathrm{thr}) = u_4 $, where $u_4$ is the specific energy that corresponds to temperature of $T_4 = 10^4\, \mathrm{K}$. This implies that the pressure is a continuous function of density and should stay constant before and after the onset of the star formation, which occurs at $\rho=\rho_\mathrm{thr}$. A further step would require considering the magnetic pressure of other non-thermal pressure components at the threshold. To be more specific, instead of taking into account only the effective specific energy of the gas, we could include the magnetic energy or alternatively consider the sum of the magnetic pressure plus the gas pressure being a continuous function of density. The magnetic pressure and energy scales with $B^2$, which scales with the density as $\rho^{4/3}$ in the flux freezing limit. This would include more complications in calculating the cold fraction at the threshold and is not crucial for the simple approach we are adopting here.
 
In \cref{eq:1densA}, we notice that the density threshold depends on the molecular fraction of the gas. Since we have already set the condition of continuous pressure on the onset of star formation to be $P_4$, i.e., the pressure that corresponds to a temperature of $10^4 \, \mathrm{K}$, it is then a natural consequence to use this pressure to calculate the molecular fraction at the threshold, which will be
\begin{equation}
     \label{eq:fmol_thr}
     f_\mathrm{mol} (P_4) = \dfrac{1}{1+ P_0/P_4}.
\end{equation}
The density threshold is now calculated as
\begin{equation}
     \label{eq:1densB}
     \rho_\mathrm{thr} = \dfrac{x_\mathrm{thr}}{(1-x_\mathrm{thr})^2} \dfrac{f_\mathrm{mol}(P_4) (\beta u_\mathrm{SN} - (1-\beta)u_c)}{f_* t_\mathrm{dyn} (\rho_\mathrm{ov}) \Lambda(\rho_\mathrm{ov},u_\mathrm{SN}/A_0)}.
\end{equation}

Another parameter that we should specify, which is a free parameter of the model, is the efficiency of the star formation process $f_*$. The efficiency of star formation is related to the gas component of the galaxy, especially the cold gas. These two quantities are correlated, as observations show, according to the Schmidt-Kennicutt relation \citep[see \cref{eq:schmidt_law}]{schmidt1959rate, kennicutt1998global}, which is a tight relation between the star formation of the galaxy and the surface density of the gas in the galaxy. In order to constrain the efficiency parameter $f_*$ we run test simulations of an isolated Milky Way type galaxy, with different $f_*$ as shown in \cref{fig:f_star_param}. In order to reproduce the observed Schmidt-Kennicutt relation, the value of the efficiency parameter should be $f_*=0.1$, i.e. star formation efficiency of $10 \%$.
 
Moreover, as far as the molecular fraction is concerned, we set the parameter $P_0$ to be $P_0/k_\mathrm{B} = 35 000\, \mathrm{Kcm^{-3}}$ according to the observations presented in \cite{blitz2006role}. The authors distinguish two different groups of galaxies, and the parameter $P_0$ changes significantly for each of them. There are three galaxies that on average they find $P_0/k_\mathrm{B} = 7700 \, \mathrm{Kcm^{-3}}$ and for the rest of the galaxies they find $P_0/k_\mathrm{B} = 43000\,  \mathrm{Kcm^{-3}}$. They attribute this difference to the content of neutral hydrogen that each galaxy contains. As calculated from observations, the mean value of the projected surface density of the neutral hydrogen may be low due to intense tidal or ram pressure stripping. However, when testing our model with the different values of this parameter
the one that follows the Schmidt-Kennicutt relation remains the $P_0$/$k_\mathrm{B} = 35 000\, \mathrm{Kcm^{-3}}$.
 
The last parameter that we examine is the temperature of cold clouds. This was assumed to be constant at $T_c = 1000\, \mathrm{K}$, but this is not close to the temperatures at which stars are formed in nature. Therefore, we examine the case of lowering the cold gas temperature and run a test simulation for $T_c = 300\, \mathrm{K}$, which is motivated by \cite{murante2015simulating}. 
The lower temperature barely affects the cold fraction and the overall star formation behaviour. The density threshold changes slightly because of its dependence on $u_c$. However, we should bear in mind that the gas at temperatures lower than $10^3-10^4\, \mathrm{K}$ needs molecular cooling for proper treatment. Since this is not included in our simulations, we will keep the temperature of the cold clouds constant to $10^3\, \mathrm{K}$. A summary of the model parameters is shown in \cref{tab:params}.
 
\begin{table}
    \centering
     \caption{Summary of the parameters used in the star formation model, where $f_*$ is the star formation efficiency, $P_0/k$ expresses the external pres-
sure of the ISM, $\beta$ is the fraction of the massive, short-lived stars that explode as core-collapse SN, and $A_0$ is the scaling factor for the evaporation factor $A$. }
    \begin{tabular}{c c c c c c}
    \hline\hline
        parameter & $f_*$ & $P_0/k$     $ [\mathrm{Kcm^{-3}}] $ & $\beta$ &  $A_0$ & $T_c$  $[\mathrm{K}]$ \\
        \hline \\
        value &  0.1 & 35000 & 0.1 &  1000 &  1000  \\
        \hline
    \end{tabular}
    \label{tab:params}
\end{table}

\subsection{Supernova Magnetic field seeding}
\label{sec:beck}
For a self-consistent model, we also aim to include the effects of star formation via the stellar feedback on the magnetic field evolution of galaxies.
To do so, we couple the star formation model with the SN magnetic seeding model of \citet{beck2013SNseeding}. This model assumes that the magnetic fields do not have a primordial origin but are deposited in the ISM from SN explosions. Here, we give a summary of the model by \citet{beck2013SNseeding} and how this is altered because we couple it to the updated star formation model discussed in the previous sections.

According to \cite{beck2013SNseeding}, an extra term is included in the induction equation to include the effect of the magnetic field seeding.
Thus, the induction equation is written as
\begin{equation}
    \label{eq:induction_seed}
    \dfrac{\partial \mathbf{B}}{\partial t} = \nabla \times (\mathbf{v} \times \mathbf{B}) + \eta \nabla^2 \mathrm{B} + \left.\dfrac{\partial \mathbf{B}}{\partial t} \right\vert_\mathrm{seed},
\end{equation}
with the seeding term
\begin{equation}
    \label{eq:seed_term}
    \left.\dfrac{\partial \mathbf{B}}{\partial t} \right\vert_\mathrm{seed} = \sqrt{N_\mathrm{SN}^\mathrm{eff}} \dfrac{B_\mathrm{inj}}{\Delta t} \mathbf{e}_B,
\end{equation}
where $\mathbf{e}_B$ is a unit vector in the direction of the seeding, $B_\mathrm{inj}$ is the amplitude of the seeding field and $N_\mathrm{SN}^\mathrm{eff}$ is a normalization constant that is connected with the number of SN explosions. The last parameter is not a free parameter of the model but is calculated directly from the star formation recipe in \textsc{gadget}. For the updated star formation model, the mass of stars, $m_*$, that are formed in each timestep $\Delta$t is calculated as
\begin{equation}
    \label{eq:m_star}
    m_* = f_* f_\mathrm{mol} m_c \dfrac{\Delta t}{t_\mathrm{dyn}}
\end{equation}
where $m_c$ is the mass of the cold clouds. The effective number of SNe explosions is given by
\begin{equation}
    \label{eq:N_sn}
    N_\mathrm{SN}^\mathrm{eff} = \alpha m_*
\end{equation}
with $\alpha$ being a parameter that specifies the number of SNe explosions per solar mass. For our case the parameter $\alpha$ is $\alpha = 0.008 \, \mathrm{M_\odot ^{-1}}$ (for a Salpeter IMF).

The total injected magnetic field for all SN explosions is given by
\begin{equation}
    \label{eq:B_inj}
    B_\mathrm{inj}^\mathrm{all} = \sqrt{N_\mathrm{SN}^\mathrm{eff}} B_\mathrm{SN} \left( \dfrac{r_\mathrm{SN}}{r_\mathrm{SB}}\right)^2 \left( \dfrac{r_\mathrm{SB}}{r_\mathrm{inj}}\right)^3
\end{equation}
where $B_\mathrm{SN}$ is the mean strength of magnetic fields in SN explosions, $r_\mathrm{SB}$ is the radius of the super-bubble of the explosion assuming spherical geometry for the remnant, $r_\mathrm{SN}$ is a typical radius of the SN remnant and $r_\mathrm{inj}$ is determined by the smoothing length of the simulation and shows the region where the super-bubbles are placed. The magnetic field seeding rate for this model is calculated as
\begin{equation}
    \dot{B}_\mathrm{seed} \simeq B_\mathrm{SN} \left( \dfrac{r_\mathrm{SN}}{r_\mathrm{SB}}\right)^2 \left( \dfrac{r_\mathrm{SB}}{r_\mathrm{inj}}\right)^3 \dfrac{\sqrt{\dot{N_\mathrm{SN}}\Delta t}}{\Delta t}.
\end{equation}
The SN radius is assumed to be $r_\mathrm{SN} = 5 \, \mathrm{pc}$, a typical value for the magnetic field strength is $B_\mathrm{SN} = 10^{-4} \, \mathrm{G}$ which is then distributed in a bubble of radius $r_\mathrm{SB} = 25 \, \mathrm{pc}$.

The magnetic field configuration added to the induction equation should be divergence-free. Therefore, a straightforward way is to assume a dipole structure of the magnetic field seed, so the seeding field changes at a rate
\begin{equation}
    \label{eq:seed_term2}
    \left.\dfrac{\partial \mathbf{B}}{\partial t} \right\vert_\mathrm{seed} = \dfrac{1}{{\mathbf{r}}^3} \left[ 3 \left(\dfrac{\partial \mathbf{m}}{\partial t}\cdot \mathbf{e}_r \right) \mathbf{e}_r -  \dfrac{\partial\mathbf{m}}{\partial t} \right]
\end{equation}
with $\mathbf{m}$ the dipole magnetic moment, $\mathbf{e}_r$ is a unit vector in the $\mathbf{r}$ direction. The time derivative of each dipole moment is written as
\begin{equation}
    \label{eq:magn_moment}
    \dfrac{\partial\mathbf{m}}{\partial t} = \sigma \dfrac{B_\mathrm{inj}^\mathrm{all}}{\Delta t} \mathbf{e}_B.
\end{equation}
Here $\mathbf{e}_B$ is the direction of the seed magnetic field, which is chosen to be in the direction of the acceleration of the particle, hence $\mathbf{e}_B = a/ \abs{a}$. The normalization constant $\sigma$ is given by
\begin{equation}
    \sigma = r_\mathrm{inj}^3 \sqrt{\dfrac{1}{2}\bar{f}^3(1+\bar{f}^3)}
\end{equation}
with $\bar{f}=r_\mathrm{soft}/r_\mathrm{inj}$ is the ratio between the softening and injection length. The parameter $\sigma$ is used to normalize the energy injected, soften the magnetic dipoles in the centre, and truncate it in the scale of the injection length.
\label{sec:SN-seeding}
\section{Numerical Method}
\label{sec:numerics}
We employ the developer's version of the Tree Smooth Particle Magnetohydrodynamics (SPMHD) code \textsc{gadget}, which is based on the well-tested code \textsc{gadget2} \citep{springel2005gadget} and incorporates an updated SPH implementation introduced in \citet{beck2016}. The code follows a Lagrangian treatment of magnetohydrodynamics and a tree algorithm for Newtonian gravity. The magnetic field implementation is described in detail in \citet{dolag2009mhd}. The aforementioned model (\cref{sec:sfr42}) for star formation and feedback is coupled to this version of the code and allows a realistic and improved treatment of the ISM in large-scale simulations. This includes star formation, the SN feedback from stars, radiative cooling and magnetic fields.

For the initial conditions, we follow the methods for a galactic model first described in \citet{hernquist1993n} in the numerical implementation that has been used in \citet{springel2005gadget}. The galactic model consists of a stellar bulge, a stellar disk, and a gas disk embedded in a dark matter halo.  The dark matter and bulge follow a mass distribution of a Hernquist profile \citep{hernquist1993n}. This system is surrounded by a hot circum-galactic medium (CGM) as first presented in \citet{moster2010can}; here, we use the implementation of \citet{Ulli} that incorporates some modifications from \citet{donnert2014} concerning the sampling mechanism of the CGM's density distribution. This follows a $\betaup$-profile, namely
\begin{equation}
    \rho (r) = \rho_0 \left( 1 + \dfrac{r^2}{r_c^2} \right)^{-3 \betaup / 2}.
\end{equation}
With $\betaup=2/3$, following \citet{mastropietro2008simulating}. In the last equation, $\rho_c = 5\cdot 10^{-26} \, \mathrm{g/cm^3}$ and $r_c = 0.33\,$kpc, as shown in \citet{Ulli}. We chose this setup in order to accurately model the smooth accretion of gas towards the galactic disk, and as we use particles to discretize the induction equation, we need a carrier medium to properly deal with the vacuum boundary conditions of the magnetic field.
This set-up is already tested and described in detail in \citet{Ulli}. Although this galactic model lacks a cosmological background, it is a useful layout in order to examine the effects of the updated star formation routines in a clean environment. In \cref{tab:HMG-parameters}, a summary of the parameters for the initial conditions used is shown.
 
The galaxy has mass resolution of $4800\, \mathrm{M_\odot}$ in the gas and stellar component and $96000\, \mathrm{M_\odot}$ in dark matter. The spatial resolution is limited by the gravitational softening, which is calculated from the recursive formula
\begin{equation}
    \centering
    \label{eq:grav_softening}
    \epsilon = \epsilon_\mathrm{old} \left( \dfrac{m}{m_\mathrm{old}} \right)^{1/3}.
\end{equation}
The reference simulation for the softening that we used to obtain the force softening of our simulations is a set of \textsc{magneticum}\footnote{http://www.magneticum.org/simulations.html} simulations \citep{hirschmann2014cosmological} as a benchmark, denoted by the subscript ``old'' in \cref{eq:grav_softening}. For the gas particles the gravitational softening it is $\epsilon_\mathrm{gas} = 10\, \mathrm{pc/h}$, for the stellar particles it is $\epsilon_\mathrm{stars} = 20\, \mathrm{pc/h}$ ,  and for the dark matter particles it is $\epsilon_\mathrm{DM} = 83\, \mathrm{pc/h}$.
\begin{table}
    \centering
    \caption{List of parameters for the initial conditions of the galaxy simulation surrounded by a gas halo.}
    \begin{tabular}{l c r}
    \hline\hline 
     Disk  Parameters &  &\\ 
    \hline
        Total mass [$10^{10}\ \mathrm{M_\odot}$]  & $M_{200}$ & 100  \\
         Virial radius & $r_{200}$ & 145 \\
         Halo concentration & $c$ & 12 \\
         Spin parameter & $\lambda$ & 0.033 \\
         Disk spin fraction & $j_d$ & 0.067 \\
         Disk mass fraction & $m_d$ & 0.067 \\
         Bulge mass fraction & $m_b$ & 0.034 \\
         Disk scale length [kpc] & $l_d$ & 2.1 \\
         Disk height [l$_d$]  & $z_0$ & 0.2 \\
         Bulge size  [l$_d$] & $l_b$ & 0.2 \\
    \hline
    \end{tabular}
    \label{tab:HMG-parameters}
\end{table}

\begin{table}
    \centering
    \label{tab:HMG-Nparticles}
    \caption{Total particle number for each particle type of the galaxy, the gas and dark-matter halo system used as initial conditions for our simulations.}
    \begin{tabular}{l c c}
    \hline\hline
    Particle Numbers [$10^6$] & & \\
    \hline
    Gas in the disk  & $N_\mathrm{gd}$ & 1.2  \\
    Stellar Disk & $N_\mathrm{sd}$ & 4.8 \\  
    Stellar Bulge & $N_b$ & 2.0 \\
    Dark matter & $N_\mathrm{DM}$ & 6.9 \\
    \hline
    \end{tabular}
\end{table}

\section{Results for a Milky Way type galaxy}
\label{sec:results}
In this section, we present the results of a Milky Way-type galaxy (High Mass Galaxy, HMG hereafter), considering two different configurations for the initial conditions of the magnetic field. In the first model, we use a primordial magnetic field of strength $10^{-9}\, \mathrm{B}$ in $\hat{x}$ direction of a Cartesian coordinate system (to be called \textit{HMG-B}), which is chosen to resemble an already partially amplified magnetic field. The second model includes the full version of the new star formation model coupled with the SN seeding prescription (to be called \textit{HMG-snB}). The magnetic seeding from SN explosions is a self-consistent way to account for the magnetic fields in the galaxy and allow the interaction with the star formation and feedback processes within the ISM. Additionally, we include a run without magnetic fields, i.e., \textit{HMG-noB} for comparison. To directly compare with SH03, we run two different simulations using the SH03 model, i.e., one without magnetic fields, to be referred to as \textit{HMG-noB-SH} and one with the full SH03 star formation and SN feedback model, to be called \textit{HMG-snB-SH}.

\subsection{General properties and morphology of the galaxy}
\begin{figure}
      \includegraphics[width=0.99\columnwidth]{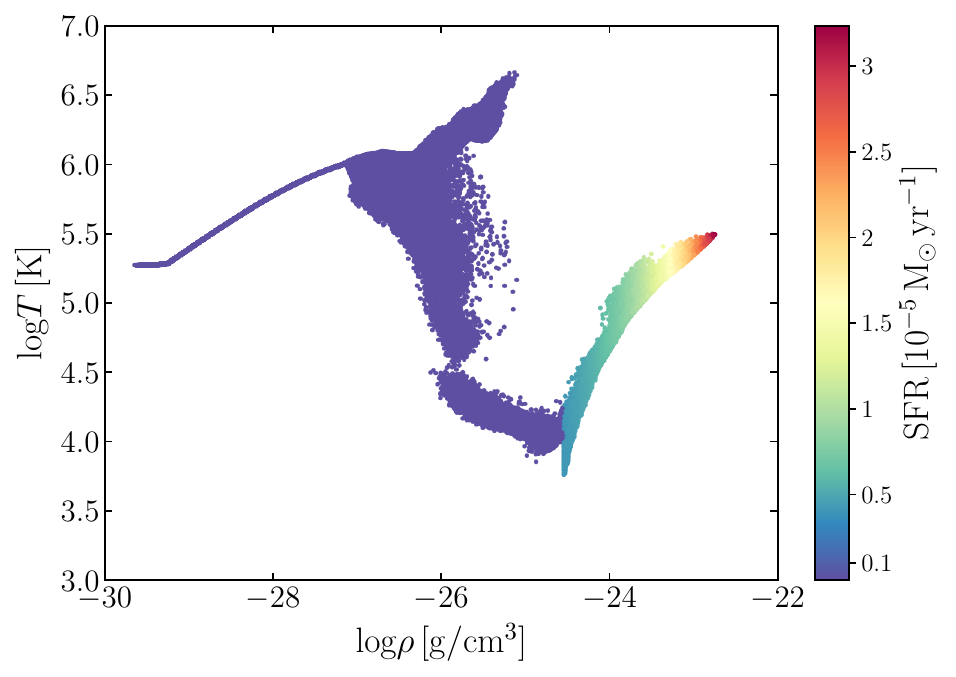}
      \includegraphics[width=0.99\columnwidth]{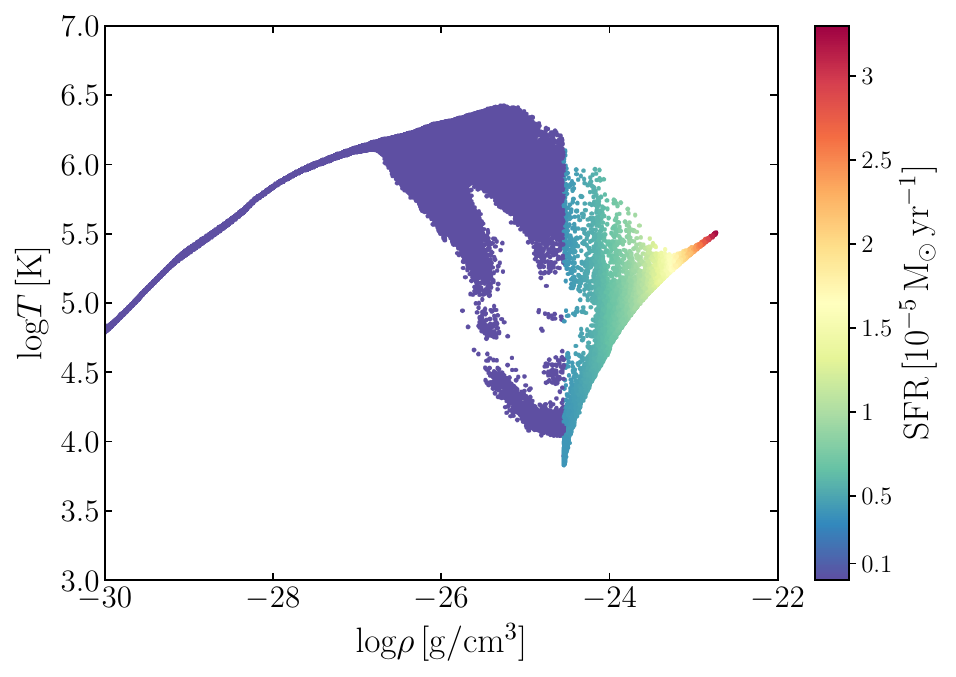}
    \caption{Temperature-density phase diagram, colour-coded by the star formation rate of each gas particle. The left upper part of the diagram corresponds to the gas halo particles, while the lower right part corresponds to the gas content of the galaxy. The top panel is at the beginning of the simulation, and the bottom is after 1\,Gyr of evolution.}
    \label{fig:phase-diagram}
\end{figure}

Before taking a closer look at the morphology of the galaxy and the structure of the magnetic field in the galaxy, it is interesting to examine the temperature-density phase diagram. In Figure \cref{fig:phase-diagram}, we show the temperature of the gas as a function of the density at the start of the simulation for the model \textit{HMG-B} for two different points in time, i.e. at the initialisation of the galaxy simulation (upper panel) and after 1\,Gyr of evolution (bottom panel). The colour bar shows the star formation rate for each computational particle. The prominent upper-left purple patch corresponds to the hot gas halo surrounding the galaxy, and it is evident that a part of it cools down as it accretes onto the galaxy. The bottom-right part of the plot corresponds to the gas disk of the galaxy. There is a clear separation based on the star formation rate, which is correlated to the density threshold imposed by our modified star formation recipe. It is worth mentioning that this density threshold is correlated with the molecular fraction we obtain directly from the total hydrostatic pressure in the galaxy's mid-plane. Moreover, we note that the temperature is interpreted as an effective temperature of the particles in the multi-phase model since the cold clouds are kept to a constant temperature of $T_c= 10^3$\,K. This is usually much lower than the temperature of the hot gas in the ISM, which easily exceeds $10^5$\,K \citep{mckee1977theory}.

%The plot in the bottom panel is shown after 1\,Gyr of evolution. 
As the galaxy interacts with the CGM, the gas halo cools down, and there is an inflow of fresh gas onto the galactic disk. A fraction of the gas is not star-forming with higher temperatures that stream from the cold disc towards the hot gas halo. This fraction of the gas is pushed out of the disc by the rising magnetic pressure, which generates an outflow of low, non-star-forming flow of gas that rises above the disc until it reaches pressure equilibrium with the surrounding hot gas of the CGM. The gas pushed out remains always below the density threshold for star formation. We find a very similar structure in the phase diagrams for the model \textit{HMG-snB}.

\begin{figure}
     \centering
     \includegraphics[width=1.0\linewidth]{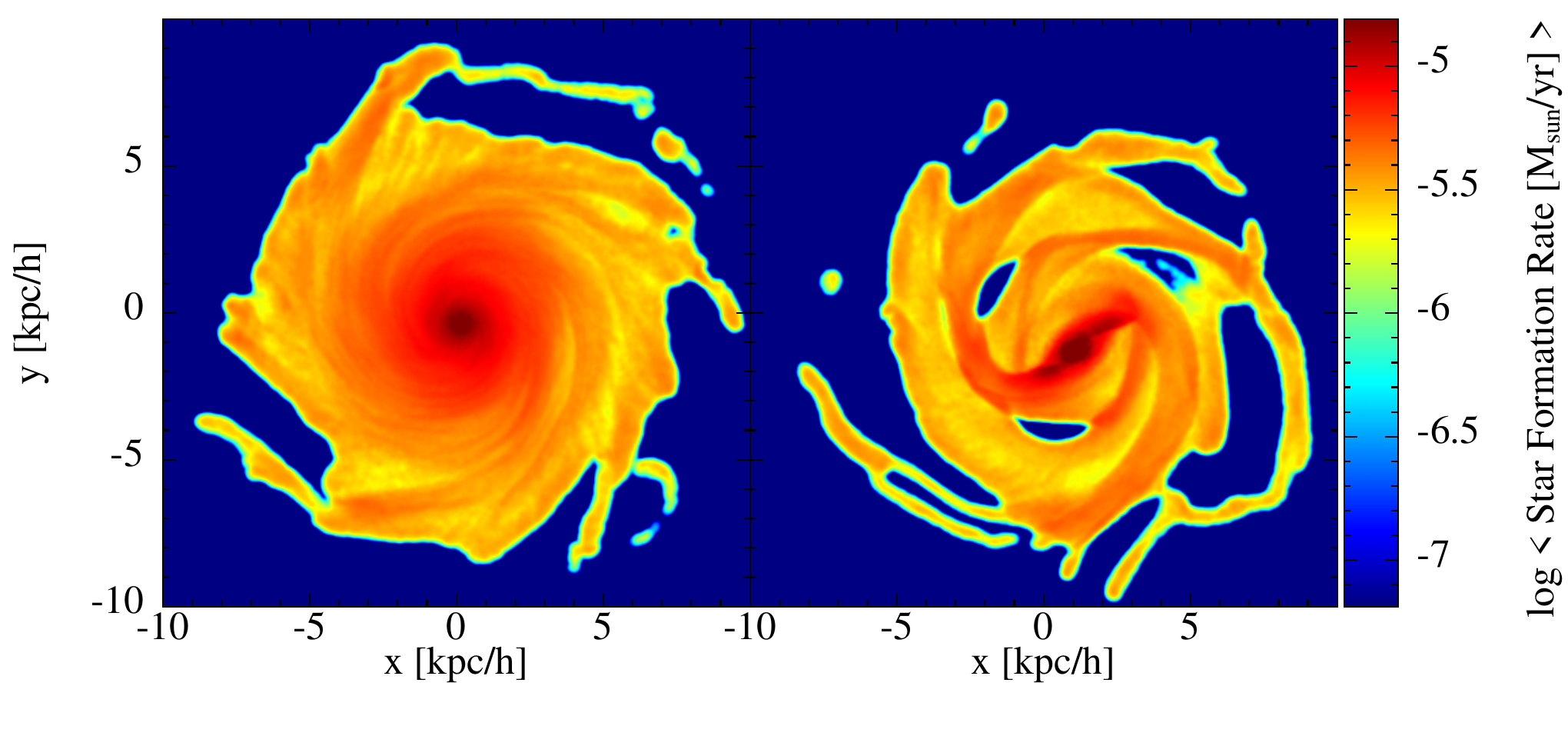}
     \caption{Projection on the xy plane 
     of the galaxy for the HMG-B simulation, colour-coded by the star formation rate. The star formation rate is integrated along the z-axis (perpendicular to the plane shown). The left panel corresponds to 1\,Gyr of evolution and the right to 2\,Gyr. A very similar structure is also found in the HMG-snB model.}
     \label{fig:slice-SFR-B}
\end{figure}{}

\begin{figure}
    \centering
    \includegraphics[width=\columnwidth]{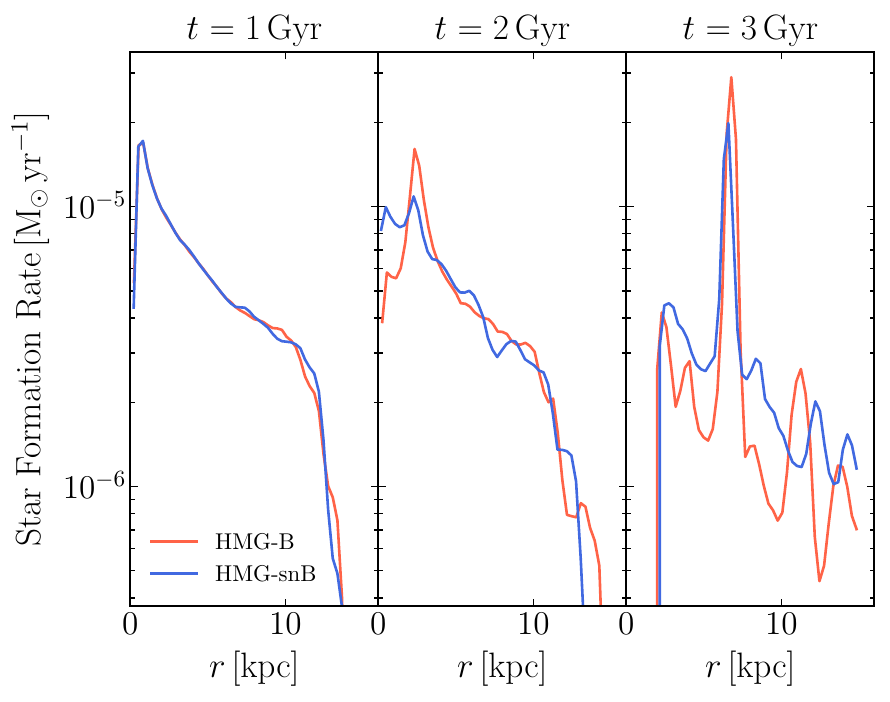}
    \caption{Radial profile of the star formation rate of the galaxy after 1, 2 and 3\,Gyr of evolution. The blue lines show the HMG-snB simulation, and the red lines show the HMG-B model. The prominent peak at 3\, Gyr of evolution is due to the outflow created; see text for details.}
    \label{fig:sfr-r}
\end{figure}

In \cref{fig:slice-SFR-B}, we show a projection of the star formation rate of the galaxy for the model \textit{HMG-B} after 1\,Gyr (left panel) and 2\,Gyr (right panel) of the evolution of the system. We find similar morphological structures for the model \textit{HMG-snB}. This can be explained by the fact that, at this time, the magnetic fields are low in both models and, therefore, are not significant enough to alter the dynamics of the galaxy. The star formation rate follows the density structure of the galaxy. Thus, our new model predicts higher star formation rates in the centre and the spiral arms, compared to the inter-arm regions and the galaxy outskirts, where the star formation rate is generally lower by a factor of 10. At the beginning of the simulation, the star formation is prominent in the spiral arms but also in the inter-arm regions. However, at later times, the star formation concentrates in the spiral arms as they are highly compressed, and the gas forms stars very efficiently over a few dynamical times. We also notice that the star formation region shrinks after 2\,Gyr of evolution (see \cref{fig:slice-SFR-B}). This can be interpreted by the fact that the star formation continuously drops and drops faster than the gas accretion from the CGM. At later times, star formation is concentrated in the centre of the galaxy in a bar-like structure fed by the hot gas from the CGM. As the CGM consists of slowly rotating low angular momentum gas, the hot material quickly assembles in the centre of the disc, triggering the formation of a bar that makes the accretion of hot gas to the centre even more efficient.
This process subsequently triggers an outflow that is generated by the magnetic pressure. We will discuss the process in more detail in \cref{sec:radial}. 

\begin{figure}
    \centering
    \includegraphics[width=\columnwidth]{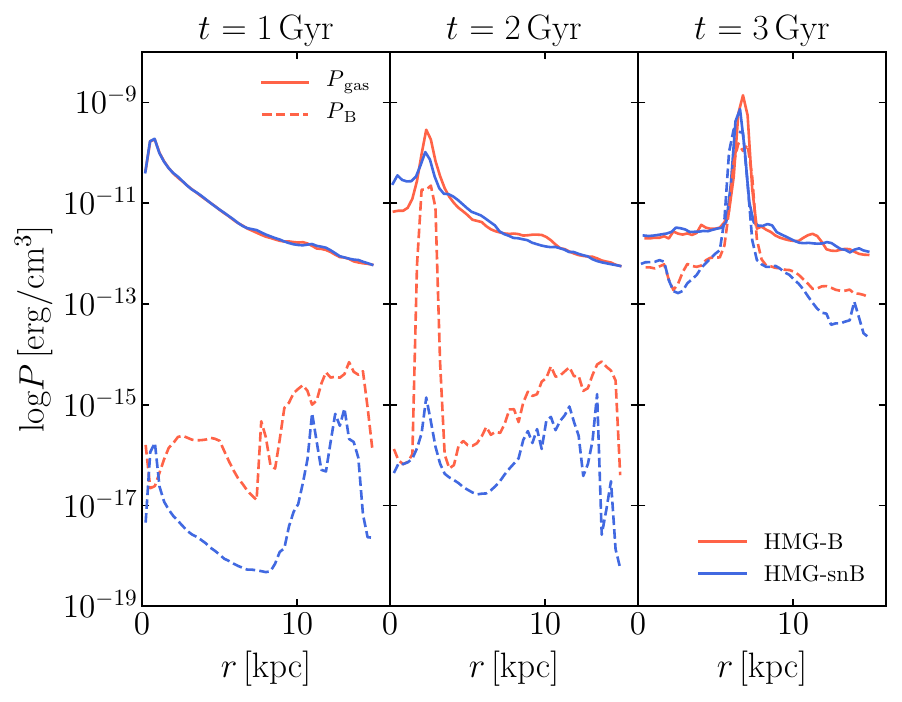}
    \caption{Radial pressure profile for 1,2 and 3\,Gyr of evolution. The blue lines show the gas pressure and the magenta lines show the magnetic field pressure. For both cases, the solid lines are the pressure components from the HMG-B simulation, while the dashed lines are from the HMG-snB.}
    \label{fig:pressure-r}
\end{figure}{}

\begin{figure}
    \centering
    \includegraphics[width=\columnwidth]{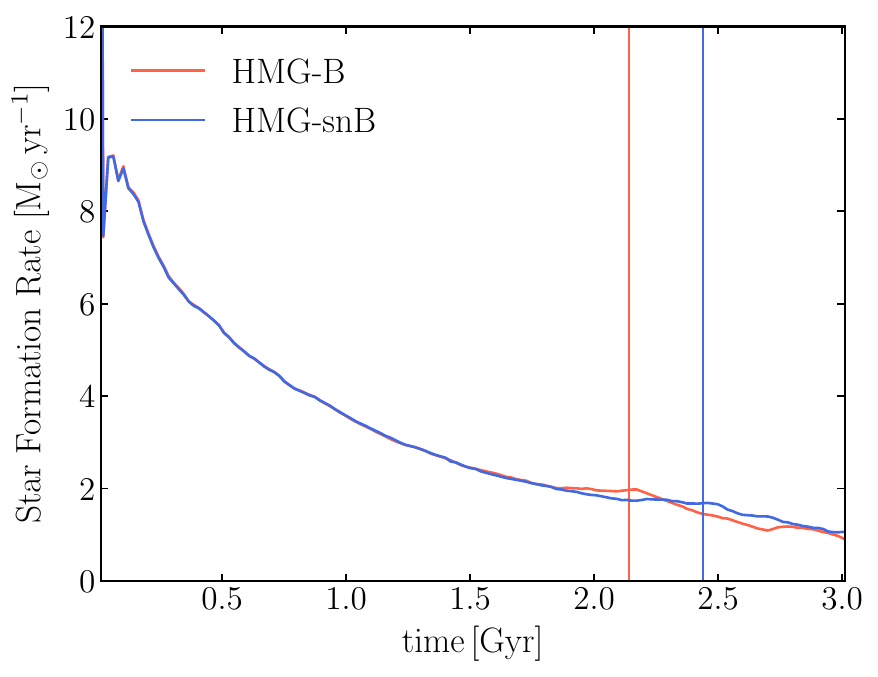}
    \caption{Total star formation rate of the galaxy as a function of time. The blue line shows the rate for the HMG-snB model and the red line for the HMG-B. The vertical lines correspond at the start of the outflow for each model following the same colour code.}
    \label{fig:SFR}
\end{figure} 

\subsection{Radial evolution}
\label{sec:radial}
To gauge the consistency of our modified star formation approach, we discuss radial profiles of the fundamental quantities directly affected by our new implementation.

In \cref{fig:sfr-r}, the star formation rate over a radius of the galaxy is shown for the two different models, i.e.\textit{HMG-B}, \textit{HMG-snB}, at three different points in time, i.e., 1\,Gyr, 2\,Gyr and 3\,Gyr. The latter refers to the final time of the simulation. At early times (t = 1\,Gyr), the star formation peaks in the centre and declines in a smooth power-law as a function of the radius. The exponential density profile of the galactic disk can easily explain this behaviour. Both models agree very well and do not lead to substantial differences in the star formation rate across the disk at early times. The sharp drop at around 12\,kpc occurs due to the boundary between the galactic disk and the ambient CGM. As low angular momentum gas is accreted towards the outskirts of the disk, it quickly cools down and forms stars in the outer parts before the star formation rate drops to zero within the CGM.

At intermediate times (t = 2\,Gyr, middle panel), we notice that the star formation rate is slightly higher in the central part of the galaxy for the \textit{HMG-B} model. In that model, the magnetic field pre-exists and thus, at 2\,Gyr, it has already been amplified enough to assist star formation, as suggested by our model.
In contrast, the \textit{HMG-snB} model starts with no magnetic field, and therefore, it needs more time to build up a field with sufficient strength. The same reason applies to the outskirts of the galaxy. In the \textit{HMG-snB} model, the star formation rate drops off in a smaller radius compared to the \textit{HMG-B}, because there is no magnetic field at initialization, leading to no contribution of the magnetic field pressure in the molecular fraction of the gas and therefore in the star formation rate. 

The right panel shows the evolution after 3\,Gyr. The star formation rate shows multiple peaks across the galaxy because prominent spiral arms are formed, as also shown in \cref{fig:slice-SFR-B}. 
%The blue line shows the star formation rate for the \textit{HMG-snB} model. 
The star formation rate for the \textit{HMG-snB} model shows different and more prominent spikes (compared to the \textit{HMG-B} model) because the star formation directly follows the magnetic field seeding process in the spiral arms, which increases the molecular fraction in those regions and subsequently that increases the star formation rate.

The radial pressure profiles of the gas and magnetic components are given in \cref{fig:pressure-r}. The (total) gas pressure is shown in solid lines, and the magnetic pressure - defined as $P_\mathrm{B} = B^2/8 \pi$ - is shown with dashed lines. The red lines refer to the \textit{HMG-B} simulation and the blue lines to the \textit{HMG-snB} simulation. Initially, the gas pressure dominates over the magnetic pressure since the timescale of magnetic field amplification on this system is large (of the order of Gyr). In both models (\textit{HMG-B} and \textit{HMG-snB}), the gas pressure has a similar behaviour with higher values in the centre and declining profile towards the outskirts of the galactic disk. The magnetic field pressure is relatively low at the beginning of the simulation (see the left panel in \cref{fig:pressure-r}) due to the low values of the magnetic field strength, but it rises in the outer parts of the galaxy because the magnetic field is amplified there faster due to rotation. In the case of \textit{HMG-snB}, the magnetic pressure is lower, compared to the one in \textit{HMG-B}, in the beginning since we start with no magnetic field and seed it gradually through the stellar feedback.
 
At 2 and 3\,Gyr, there is a pronounced peak in the magnetic and gas pressure. During the evolution of the galaxy, strong magnetic fields are built up in the centre of the galaxy. This results in a magnetically driven outflow that has been extensively discussed in \cite{Ulli} and \cite{Ulli-Bwinds}. This peak in the pressure components explains the peak in the star formation profiles as our model predicts (see \cref{eq:1fmol}). 

\subsection{Evolution of the star formation rate}
The total star formation rate over time is shown in \cref{fig:SFR}. The star formation peaks (starburst) at the beginning of the simulation because of the high cold fraction at the initial conditions (same as SH03). A fraction of the gas is kept in the cold star-forming phase, but there is no stellar feedback yet in the form of SNe to heat the cold gas. This results in the initial starburst. This can be prevented by steering ISM-turbulence at the beginning of the simulation. However, this is deliberately not done as we want to compare as closely as possible to the results of SH03. The star formation rate gradually decreases with time since the star formation consumes a large part of the gas. After almost stabilising the star formation rate, at 2.1\,Gyr, the \textit{HMG-B} simulation indicates a sudden drop. This can be explained by the rising magnetic pressure in the centre, which subsequently drives an outflow that is correlated to adiabatic compression of the magnetic field due to a bar-formation process. This increases the magnetic pressure in the centre to a similar order to the magnitude of the thermal pressure (see middle and right panel of \cref{fig:pressure-r}; radial pressure profiles), leading to an outflow driven by the magnetic pressure (see \citet{Ulli-Bwinds} for details).

We note that we find a difference for the model \textit{HMG-SnB} compared to the previously obtained results from \citet{Ulli}. While \citet{Ulli} reported that there is no difference in the magnetically driven outflow between the models \textit{HMG-B} and \textit{HMG-snB}, we find that the outflow is delayed by 300\,Myr in the \textit{HMG-snB} model compared to the \textit{HMG-B} model. This is driven by the modified star formation recipe. In the model \textit{HMG-snB}, there is no primordial magnetic field; thus, the field amplification requires longer to build a strong enough field to drive the outflow. Additionally, the modification in the star formation model changes the timing of the episodes of star formation and leads to an overall slightly different evolution.

After $\sim$3\,Gyr or evolution, the star formation rate for both models reaches an asymptotic value of $\sim$1\,$\mathrm{M_\odot / yr}$, which is in line, within uncertainties, with observations of the star formation rate of the Milky Way \citep[e.g.,][]{robitaille2010present,Elia2022,Zari2023}.

\subsection{Comparison with SH03}
\begin{figure}
     \centering
     \includegraphics[width=\columnwidth]{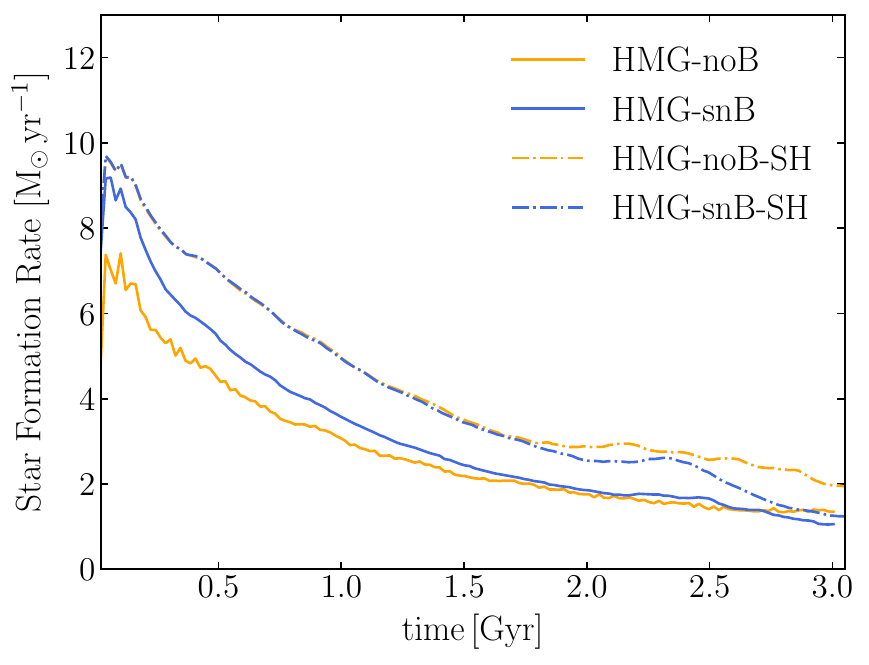}
     \caption{Star formation rate over time for the model HMG-snB, the same simulation run without magnetic fields (HMG-noB) and the simulations mentioned above using the SH03 model for star formation and feedback.}
     \label{fig:sfr-comparison}
\end{figure}
In \cref{fig:sfr-comparison}, we show the star formation rate over time for the model \textit{HMG-snB}, the same simulation run without magnetic fields (\textit{HMG-noB}) and their counterparts using the SH03 model. All simulations use the same initial conditions described in \cref{sec:numerics}. As already mentioned, the peak of star formation rate at the beginning of each simulation is due to the fact that initially, a lot of gas is initiated in the cold phase, which is star-forming. During the first Gyr of evolution, the star formation rate remains lower for our model compared to the star formation rate for the SH03 model. This can be explained by the fact that our model includes the molecular phase of the ISM, which gives rise to star formation.
In the simulations without magnetic fields (orange lines, i.e., HMG-noB and HMG-noB-SH), the star formation rate stays consistently lower due to the reasons mentioned above. The simulations, including magnetic fields (via SN seeding), reach the same asymptotic value after 3\,Gyr of evolution. We also notice that in our model (HMG-snB), there is a continuous decrease of the star formation rate, in contrast to the sudden drop noticed at $\sim$2.4\,Gyr in the model with the SH03 star formation recipe, which is an impact of the magnetically driven outflow \citep{Ulli-Bwinds}. 

\subsection{Magnetic Fields}
\label{sec:MagneticFields}
\begin{figure*}
    \centering
    \includegraphics[width=0.8\textwidth]{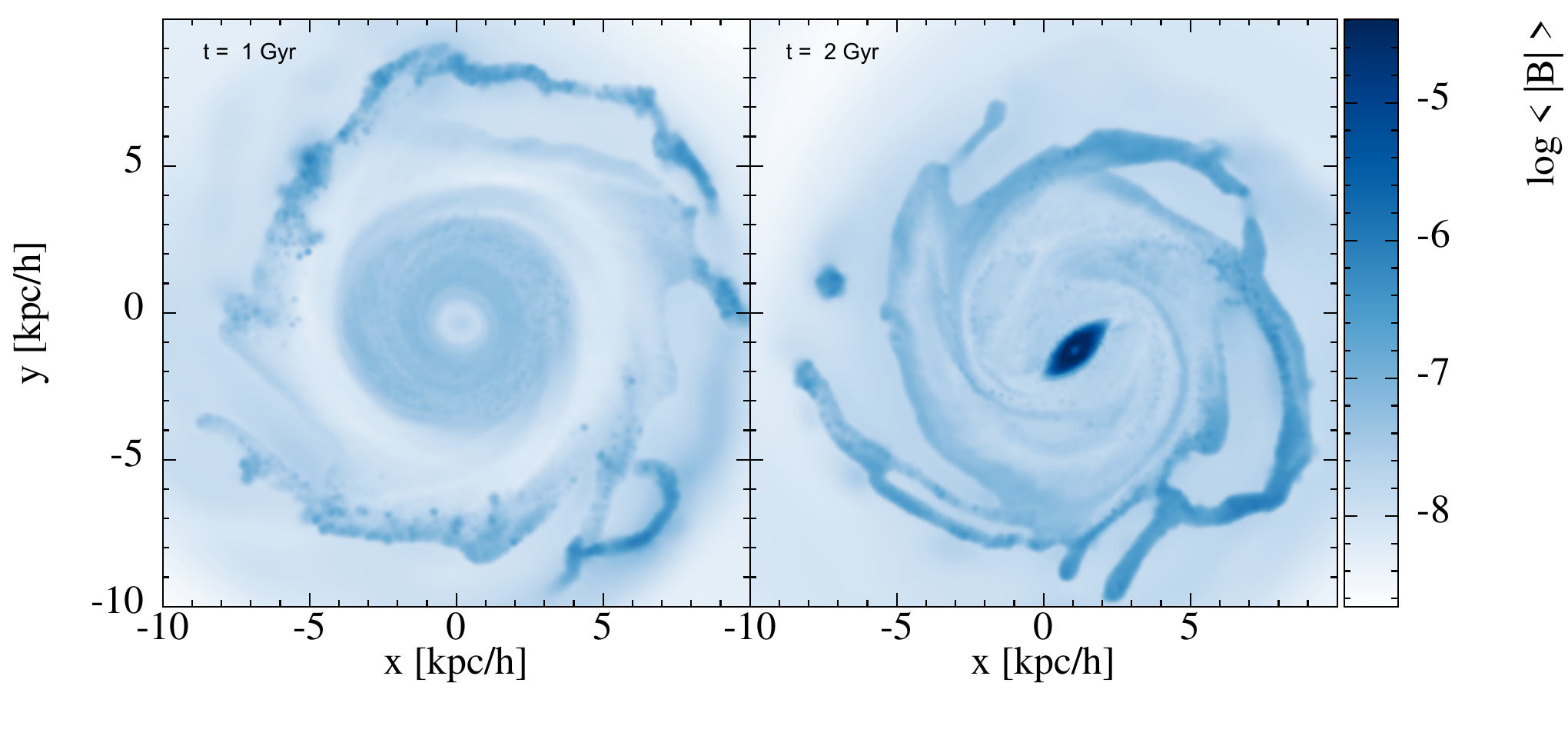}
    \includegraphics[width=0.8\textwidth]{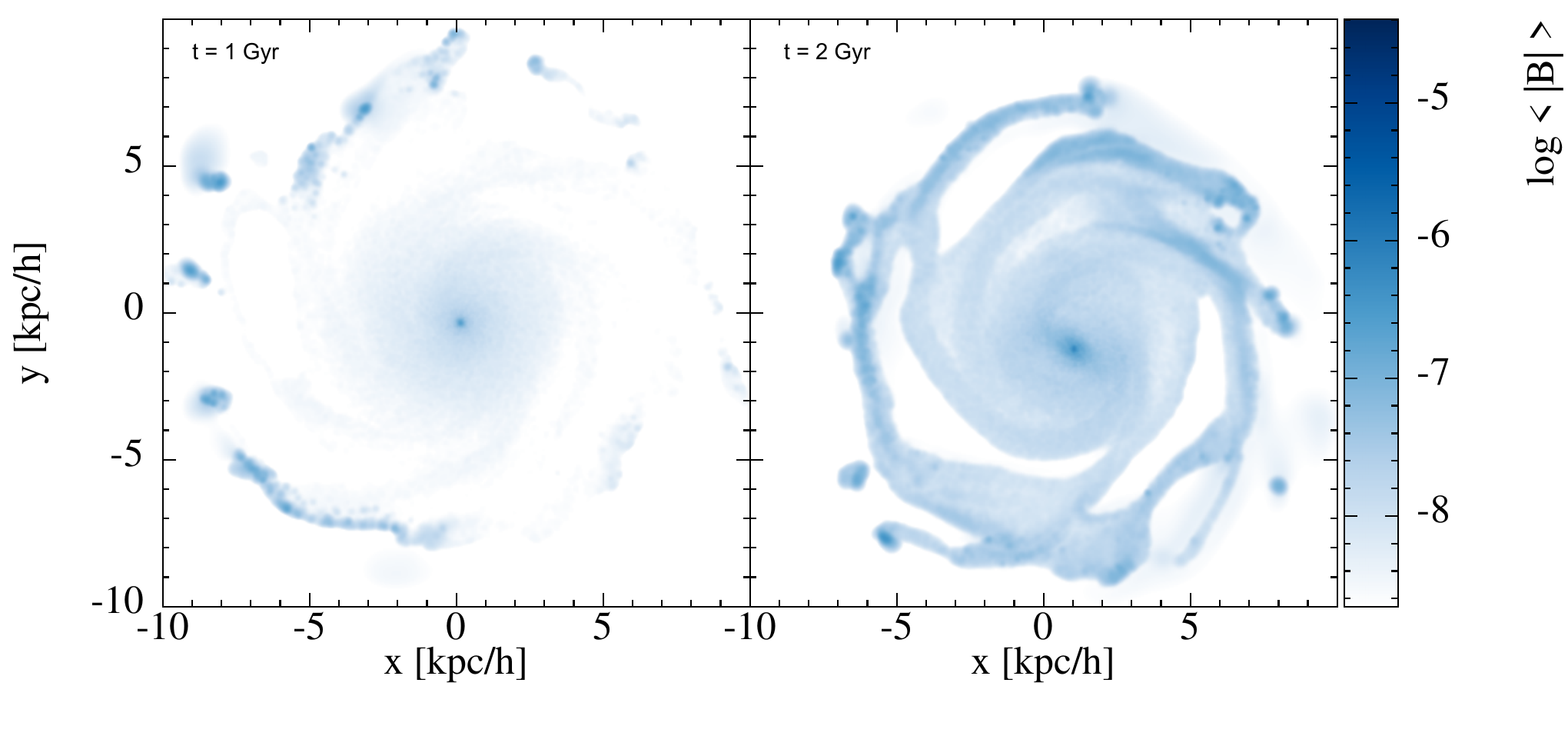}
    \caption{Slice of the xy galaxy plane of the HMG model with two initial magnetic field configurations. The top panels show the simulation with the primordial initial magnetic field, i.e., model HMG-B, at 1\,Gyr (left) and 2\,Gyr (right). The bottom panels show the simulation with the SN seeding magnetic field, i.e., model HMG-snB, at 1\,Gyr (left) and 2\,Gyr (right). The colour code indicates the magnetic field strength on a logarithmic scale.}
    \label{fig:Bsplash}
\end{figure*}

\begin{figure}
    \centering
    \includegraphics[width=\columnwidth]{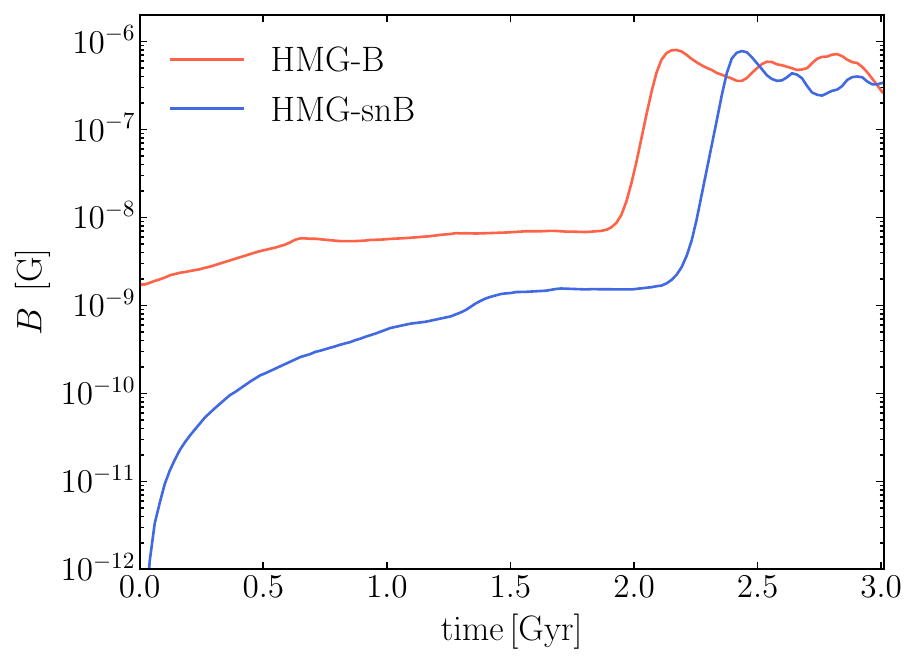}
    \caption{Total magnetic field strength of the galactic field as a function of time. The dark blue corresponds to the simulation with the primordial magnetic field, while the light blue corresponds to the one with the SN seeding magnetic field.}
    \label{fig:Btime}
\end{figure}
\begin{figure*}
    \centering
    \includegraphics[clip,trim={0cm 0cm 5cm 0cm},width=0.95\linewidth]{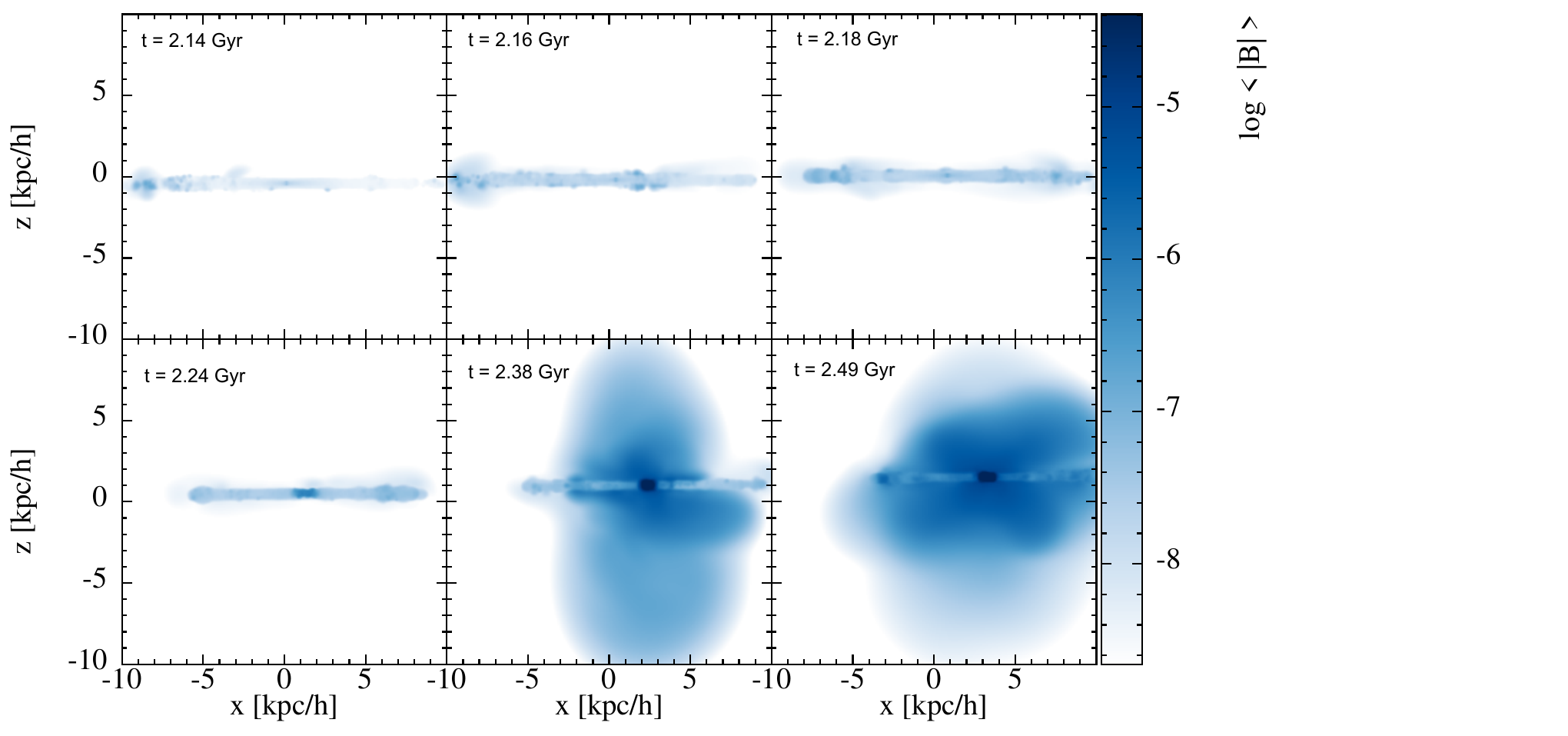}
    \caption{Edge on cross-section slices of the simulation HMG-B that show the start and evolution of the magnetically-driven outflow. The outflow starts at 2.14 Gyr, and we show how it is manifested until 2.49 Gyr. The colour bar (same scale in every plot) shows the magnetic field strength in Gauss.}
    \label{fig:HMG-outflow-B}
\end{figure*}
In the discussed simulations, we compare two different initialisation configurations for the magnetic fields on the galactic system. The model \textit{HMG-B}, which is initialised with the primordial magnetic field of $10^{-9}\, \mathrm{G}$ and the model \textit{HMG-snB}, which is based on magnetic seeding from SN explosions (see the introductory part in \cref{sec:results}).
 
\Cref{fig:Bsplash} shows slices of the galactic plane for the \textit{HMG-B} (top panels) and the \textit{HMG-snB} (bottom panels) for two different times of evolution, i.e. 1\,Gyr (left) and 2\,Gyr (right). The colour code shows the magnetic field strength for each particle of the computational domain. In the case of the \textit{HMG-B} simulation, the magnetic field is strong enough from the beginning, while the \textit{HMG-snB} needs more time to build up a field of similar strength. Due to the rotation of the galactic disk, the magnetic field is amplified through the already known $\Omega$ mechanism as a part of the $\alpha \Omega$ dynamo \citep{ruzmaikin1988magnetism}. In the bottom panels, it is evident that the timescale for the amplification of the $\mathbf{B}$ in the outskirts of the galactic plane is shorter compared to the amplification in the central parts due to turbulent motions. The galaxy (in the textit{HMG-snB} model) needs substantial time to produce stars, from which a few of them will end their lives in SN explosions and, in turn, enrich the ISM with magnetic fields and hot gas.

The total magnetic field strength evolution across time is shown in \cref{fig:Btime}. The red line corresponds to the \textit{HMG-B} simulation, and the blue line corresponds to the \textit{HMG-snB}. Initially, in the \textit{HMG-B} model, the primordial magnetic field configuration is amplified in just an order of magnitude and stays constant until it is significantly boosted at $\sim$ 2\,Gyr for the \textit{HMG-B} model. The model in which the magnetic fields are initiated with the SN seeding model (\textit{HMG-snB}) shows the steep rise of the magnetic field strength followed by an exponential growth, which is characteristic of these systems \citep{pakmor2013simulations,Ulli,Ntormousi2020,Pakmor2024}. The sudden magnetic field boost is also shown in the \textit{HMG-snB} simulation, although it appears later in time ($\sim$ 2.2 Gyr). In both models, the magnetic field strength saturates of orders of $\mu G$, which is also what is measured in observations of spiral galaxies \citep[e.g.,][]{chyzy2008magnetic,Krause2019,Heesen2023}.

In \cref{fig:HMG-outflow-B}, the evolution of the magnetic field around 2.1\,Gyr is shown in a series of edge-on cross-section slices of the galaxy for the \textit{HMG-B} model. As discussed above, the magnetic field is amplified in the central part of the galaxy, and it is correlated with a peak in the thermal gas and magnetic pressure and thus star formation rate (last panels of \cref{fig:sfr-r} and \cref{fig:pressure-r}). Due to this steep pressure gradient, the low-density but high magnetic field strength outflow is driven. The outflow is numerically manifested as SPMHD particles ejected from the disk towards the gas halo, contributing to the gas halo's magnetic field. These particles act as carriers for the magnetic field, allowing the outflow to evolve. The geometrical difference between our simulations and those of \citet{Ulli} is that the outflow in our case is not accelerated to high velocities and thus remains closer to the galaxy (max $v_z \sim$ 50\,km/s).
The results of the simulation with the SN-seeded magnetic fields (i.e., \textit{HMG-snB} model) are similar apart from the time delay in the outflow. Our results are also in accordance with \citet{pakmor2013simulations}, who found low-density but highly magnetised rising bubbles around the galaxy at $\sim$ 2\,Gyr with similar geometry. These outflows are also observed in different simulations, but in other cases, they are claimed to have a different origin, for instance, in \citet{marinacci2011}, where SN-powered bubbles drive the outflows.
 
\subsection{Correlation of the magnetic field and the star formation surface density}
\label{sec:B-SFR}
\begin{figure}
    \centering
    \includegraphics[width=\columnwidth]{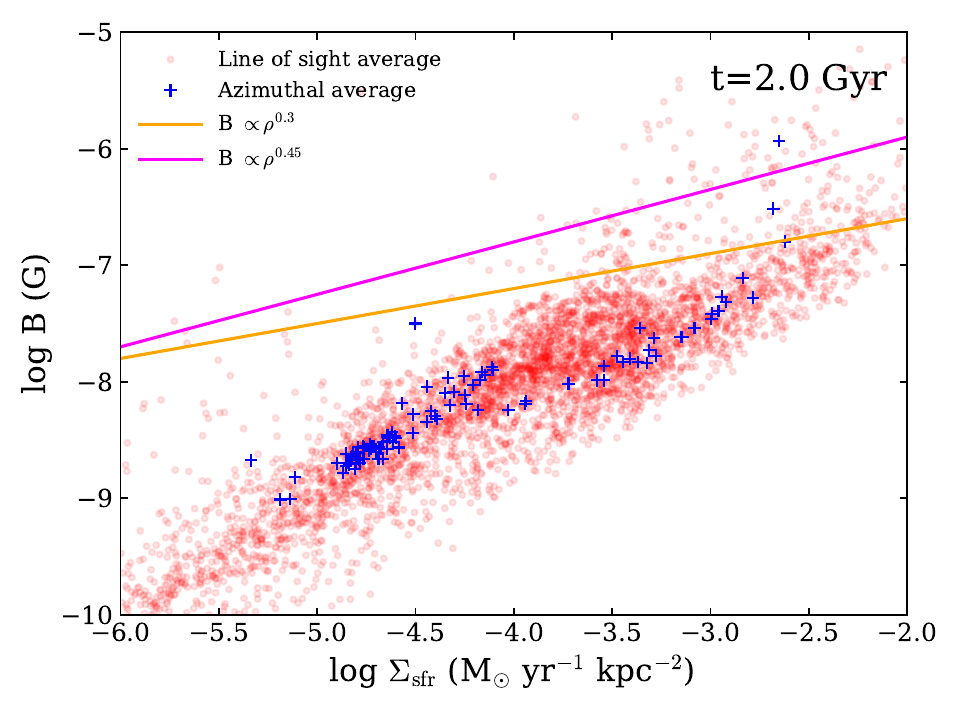}
    \caption{Logarithm of magnetic field strength over the surface density of the star formation rate in the galactic disk. Red symbols show the line of sight averaged values and blue crosses the azimuthal average. The two lines represent theoretical scaling \citep{Schleicher2013} between magnetic field strength and star formation surface density in molecular gas (magenta line) and neutral gas (orange line). The data are calculated after 2 Gyr of evolution for the HMG-snB model.}
    \label{fig:B-gas}
\end{figure}

We investigate the scaling of the global magnetic field with the surface density of the star formation rate in our simulated galaxy with the full star formation and feedback model (model \textit{HMG-snB}). Theoretical studies \citep[e.g.,][]{Schleicher2013} have shown that the magnetic field correlates with the surface density of the star formation rate as $\mathrm{B} \sim \Sigma^{1/3}_\mathrm{SFR}$. 

In \cref{fig:B-gas} the magnetic field strength over the surface density of the star formation rate is shown in logarithmic scaling. The red dots are averages in the line of sight across the galactic disk, and the blue crosses are azimuthal averages. We also plot the expected theoretical scaling \citep{Schleicher2013}, namely, the magnetic field relation with the surface density of the neutral gas (orange line) and the magnetic field with the surface density of the star formation rate coming from the molecular fraction. Our results are in very good agreement with the predictions, which are also verified by observations \citep[e.g.,][]{Niklas1997,Krause2015BGalaxy,Manna2023B-gas}, which find that the total magnetic field correlates with the gas density as $B_\mathrm{tot} \sim \rho^{0.5}$. It is interesting to notice that our model tends to follow the prediction of the magenta line more closely, i.e., the correlation between the global magnetic field and the $\Sigma_\mathrm{SFR}$ from molecular gas. This reflects directly on the skeleton of our model since the star formation is driven by the molecular gas of the ISM. 
 
\subsection{Comparison to observations}
\begin{figure*}
     \centering
     \includegraphics[clip,trim={5.5cm 0cm 0cm 0cm},width=\textwidth]{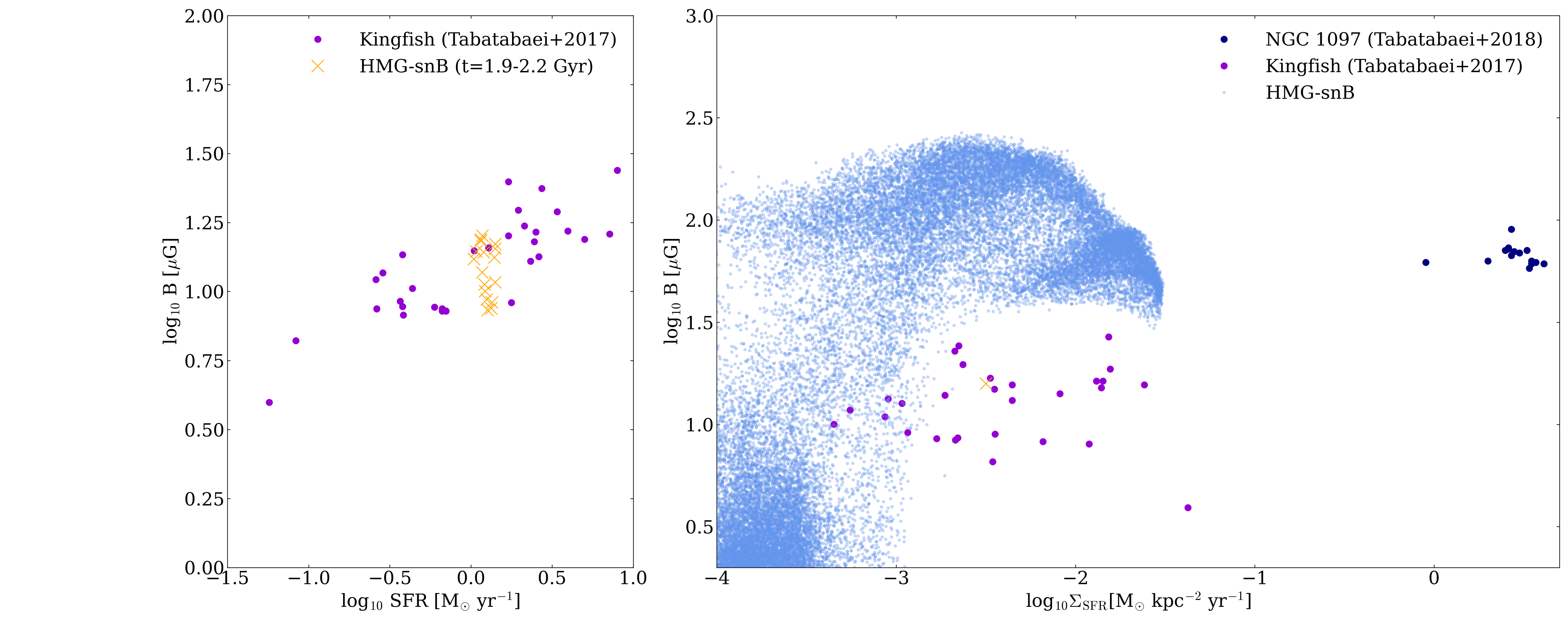}
     \caption{Logarithm of the magnetic field strength as a function of the star formation rate (left panel). Observational data are the galaxies from the KINGFISH sample \citep{2017ApJ...836..185T}, blue dots are the model HMG-snB for times between 1.9 and 2.2\,Gyr. The right plot shows a comparison of particle base at 2.2\, Gyr, highlighting the negative correlation at a very high star-formation rate. This is qualitatively in line with observation in NGC 1097 \citep{2018NatAs...2...83T}, although our simulations do not resolve the very high star-formation density as in the observed giant molecular clouds. In addition, the purple data points are the KINGFISH sample, divided by the optical size, to obtain a star-formation density in the spirit of what was done in \citet{2017ApJ...836..185T}. The orange cross (right panel) is what we get for model HMG-snB when we divide the total star-formation rate by the size of the simulated galaxy.}
     \label{fig:local-sfr-comparison}
 \end{figure*}
To compute the observed mean magnetic field from the simulated galaxy (model \textit{HMG-snB}), we computed the averaged magnetic field within the stellar body of the galaxy, i.e., taking a cylinder with a radius of $10\, \mathrm{kpc}$ and a height of $\pm0.5\, \mathrm{kpc}$. The mean magnetic field within our simulated galaxy thereby compares very well with observations, as seen in the left panel of \cref{fig:local-sfr-comparison}, where we plot the mean magnetic field as a function of global star-formation rate. Complementary to \cref{sec:B-SFR} and \cref{fig:B-gas}, it is shown here how the global magnetic field of the galaxy correlates with the surface density of star-forming gas.

To confirm the observational indication that locally, in highly star-forming molecular clouds, the local star-formation density is anti-correlated with the local magnetic fields, we computed an equivalent star-formation surface density for each of the star-forming SPMHD particles by dividing the star-formation rate of each particle by the 2D area corresponding to its volume. A comparison of this equivalent star-formation surface density with the local magnetic field carried by the SPMHD particle is shown in the right panel of \cref{fig:local-sfr-comparison}, where we also added the measurements from NGC 1097 \citep{2018NatAs...2...83T}. It is clear to see that the simulation also shows a clear anti-correlation between the local magnetic field and the local star formation rate at high star-formation rates. Even though the resolution of the simulation neither allows to resolve of individual giant molecular clouds nor the simulated galaxy to have such strong star formation as the nuclear ring in NGC 1097, it is visible that magnetic fields significantly influence and regulate the star formation. In the spirit of \citet{2017ApJ...836..185T}, where they computed a global star-formation rate density by dividing through the optical size of the galaxy, we added the KINGFISH sample (purple data-points in our \cref{fig:local-sfr-comparison}, see Figure 8 in \citet{2017ApJ...836..185T}) as well as the value obtained for our simulated galaxy (orange cross in right panel of \cref{fig:local-sfr-comparison}), again showing very nice agreement.

\section{Summary and Conclusions}
We present an updated model for star formation and feedback in the cosmological SPMHD code \textsc{gadget} that accounts for the magnetic field effects on the star formation process in galaxies. We modify the model equations of the widely used sub-grid model for star formation and feedback by \citet{springel2003cosmological} to include non-thermal components of the ISM in the modelling of unresolved scales. The novel feature of our model is that we include an extra component of the multi-phase ISM apart from the hot and the cold phase, i.e., the molecular component from which the star formation arises. Using an observationally derived relation between the fraction of the molecular gas in a galaxy and the total pressure in the galaxy, we include the effects of magnetic fields via the magnetic pressure of the galaxy as a regulation of the star formation process. This pressure-oriented approach allows the inclusion of non-thermal components of the ISM in the star formation process. In addition, our model is coupled to the magnetic SN seeding model (see \cref{sec:SN-seeding}), which deposits the magnetic field in a galaxy via SN explosions and does not assume the primordial origin of the magnetic field in galaxies. The effective number of SN explosions, that determine the magnetic field input, is directly calculated from our model. Thus, our model incorporates a self-consistent interplay between magnetic fields and star formation in galaxies.

We implemented the model in our developer's version code \textsc{gadget} and tested it by computing an isolated Milky Way-type galaxy embedded in a hot CGM environment. Our results for the magnetic field growth and star formation rate over time agree with previous studies of such simulations. Apart from that, our model is able to predict both local and global properties of the magnetic field configuration within the galaxy. Our main conclusions are summarized as follows: 
\begin{itemize}
    \item Our new star formation recipe, tested in an idealized galactic environment, reproduces the basic galactic properties compared to previous studies \citep[e.g.,][]{pakmor2013simulations,Ulli,Ntormousi2020,Wibking2023,Pakmor2024}. Galactic magnetic fields are amplified and reach observable values of $\mu$G. When the magnetic field is sufficiently amplified in the galactic centre, it can drive highly magnetised and low-density outflows, enriching the CGM with magnetic fields.
    \item We find that the total magnetic field of the galaxy scales with the surface density of the star formation rate, which is directly related to the gas density of the galaxy. Our results show good agreement with theoretical predictions \citep[e.g.,][]{Schleicher2013} and observations \citep[e.g.,][]{2017ApJ...836..185T}.
    \item We observe that the local magnetic field strength is anti-correlated with the star formation rate in the high star-formation rate regime. This indicates that strong magnetic fields can quench star formation in giant molecular clouds, as suggested by observations \cite{2018NatAs...2...83T}.
\end{itemize}

Our pressure-based star formation and feedback model can be further used to include different components of non-thermal pressure of the ISM in order to consider all the significant processes affecting star formation. Future work includes verifying the model in simulations of galaxies with a cosmological background and in large-scale cosmological simulations.

\section*{Acknowledgements}

UPS is supported by a Flatiron Research Fellowship at the Center for Computational Astrophysics of the Flatiron Institute. The Flatiron Institute is supported by the Simons Foundation. KD acknowledges support by the COMPLEX project from the European Research Council (ERC) under the European Union’s Horizon 2020 research and innovation program grant agreement ERC-2019-AdG 882679 as well as support by the Deutsche Forschungsgemeinschaft (DFG, German Research Foundation) under Germany’s Excellence Strategy - EXC-2094 - 390783311. MV is supported by the Fondazione ICSC National Recovery and Resilience Plan (PNRR), Project ID CN-00000013 "Italian Research Center on High-Performance Computing, Big Data and Quantum Computing" funded by MUR - Next Generation EU. MV also acknowledges partial financial support from the INFN Indark Grant, and by the Alexander von Humboldt Stiftung during the initial phase of this work.

\section*{Data availability}
The data underlying this article will be shared on reasonable request
to the corresponding author.

%%%%%%%%%%%%%%%%%%%%%%%%%%%%%%%%%%%%%%%%%%%%%%%%%%



\bibliographystyle{mnras}
\bibliography{references.bib} 

\begin{thebibliography}{}
\makeatletter
\relax
\def\mn@urlcharsother{\let\do\@makeother \do\$\do\&\do\#\do\^\do\_\do\%\do\~}
\def\mn@doi{\begingroup\mn@urlcharsother \@ifnextchar [ {\mn@doi@}
  {\mn@doi@[]}}
\def\mn@doi@[#1]#2{\def\@tempa{#1}\ifx\@tempa\@empty \href
  {http://dx.doi.org/#2} {doi:#2}\else \href {http://dx.doi.org/#2} {#1}\fi
  \endgroup}
\def\mn@eprint#1#2{\mn@eprint@#1:#2::\@nil}
\def\mn@eprint@arXiv#1{\href {http://arxiv.org/abs/#1} {{\tt arXiv:#1}}}
\def\mn@eprint@dblp#1{\href {http://dblp.uni-trier.de/rec/bibtex/#1.xml}
  {dblp:#1}}
\def\mn@eprint@#1:#2:#3:#4\@nil{\def\@tempa {#1}\def\@tempb {#2}\def\@tempc
  {#3}\ifx \@tempc \@empty \let \@tempc \@tempb \let \@tempb \@tempa \fi \ifx
  \@tempb \@empty \def\@tempb {arXiv}\fi \@ifundefined
  {mn@eprint@\@tempb}{\@tempb:\@tempc}{\expandafter \expandafter \csname
  mn@eprint@\@tempb\endcsname \expandafter{\@tempc}}}

\bibitem[\protect\citeauthoryear{{Agertz} \& {Kravtsov}}{{Agertz} \&
  {Kravtsov}}{2016}]{agertz2016impact}
{Agertz} O.,  {Kravtsov} A.~V.,  2016, \mn@doi [\apj]
  {10.3847/0004-637X/824/2/79}, \href
  {https://ui.adsabs.harvard.edu/abs/2016ApJ...824...79A} {824, 79}

\bibitem[\protect\citeauthoryear{{Aumer}, {White}, {Naab}  \&
  {Scannapieco}}{{Aumer} et~al.}{2013}]{aumer2013towards}
{Aumer} M.,  {White} S. D.~M.,  {Naab} T.,   {Scannapieco} C.,  2013, \mn@doi
  [\mnras] {10.1093/mnras/stt1230}, \href
  {https://ui.adsabs.harvard.edu/abs/2013MNRAS.434.3142A} {434, 3142}

\bibitem[\protect\citeauthoryear{{Basu} \& {Roy}}{{Basu} \&
  {Roy}}{2013}]{Basu2013}
{Basu} A.,  {Roy} S.,  2013, \mn@doi [\mnras] {10.1093/mnras/stt845}, \href
  {https://ui.adsabs.harvard.edu/abs/2013MNRAS.433.1675B} {433, 1675}

\bibitem[\protect\citeauthoryear{{Beck}}{{Beck}}{2015}]{BeckReview}
{Beck} R.,  2015, \mn@doi [\aapr] {10.1007/s00159-015-0084-4}, \href
  {https://ui.adsabs.harvard.edu/abs/2015A&ARv..24....4B} {24, 4}

\bibitem[\protect\citeauthoryear{{Beck} \& {Wielebinski}}{{Beck} \&
  {Wielebinski}}{2013}]{2013BeckbookObs}
{Beck} R.,  {Wielebinski} R.,  2013, {Magnetic Fields in Galaxies}.
p.~641, \mn@doi{10.1007/978-94-007-5612-0_13}

\bibitem[\protect\citeauthoryear{{Beck}, {Lesch}, {Dolag}, {Kotarba}, {Geng}
  \& {Stasyszyn}}{{Beck} et~al.}{2012}]{Beck2012}
{Beck} A.~M.,  {Lesch} H.,  {Dolag} K.,  {Kotarba} H.,  {Geng} A.,
  {Stasyszyn} F.~A.,  2012, \mn@doi [\mnras]
  {10.1111/j.1365-2966.2012.20759.x}, \href
  {https://ui.adsabs.harvard.edu/abs/2012MNRAS.422.2152B} {422, 2152}

\bibitem[\protect\citeauthoryear{{Beck}, {Dolag}, {Lesch}  \&
  {Kronberg}}{{Beck} et~al.}{2013}]{beck2013SNseeding}
{Beck} A.~M.,  {Dolag} K.,  {Lesch} H.,   {Kronberg} P.~P.,  2013, \mn@doi
  [\mnras] {10.1093/mnras/stt1549}, \href
  {https://ui.adsabs.harvard.edu/abs/2013MNRAS.435.3575B} {435, 3575}

\bibitem[\protect\citeauthoryear{{Beck} et~al.,}{{Beck}
  et~al.}{2016}]{beck2016}
{Beck} A.~M.,  et~al., 2016, \mn@doi [\mnras] {10.1093/mnras/stv2443}, \href
  {https://ui.adsabs.harvard.edu/abs/2016MNRAS.455.2110B} {455, 2110}

\bibitem[\protect\citeauthoryear{{Bieri}, {Naab}, {Geen}, {Coles}, {Pakmor}  \&
  {Walch}}{{Bieri} et~al.}{2023}]{Bieri2023}
{Bieri} R.,  {Naab} T.,  {Geen} S.,  {Coles} J.~P.,  {Pakmor} R.,   {Walch} S.,
   2023, \mn@doi [\mnras] {10.1093/mnras/stad1710}, \href
  {https://ui.adsabs.harvard.edu/abs/2023MNRAS.523.6336B} {523, 6336}

\bibitem[\protect\citeauthoryear{{Blitz} \& {Rosolowsky}}{{Blitz} \&
  {Rosolowsky}}{2006}]{blitz2006role}
{Blitz} L.,  {Rosolowsky} E.,  2006, \mn@doi [\apj] {10.1086/505417}, \href
  {https://ui.adsabs.harvard.edu/abs/2006ApJ...650..933B} {650, 933}

\bibitem[\protect\citeauthoryear{{Butsky}, {Zrake}, {Kim}, {Yang}  \&
  {Abel}}{{Butsky} et~al.}{2017}]{Butsky2017}
{Butsky} I.,  {Zrake} J.,  {Kim} J.-h.,  {Yang} H.-I.,   {Abel} T.,  2017,
  \mn@doi [\apj] {10.3847/1538-4357/aa799f}, \href
  {https://ui.adsabs.harvard.edu/abs/2017ApJ...843..113B} {843, 113}

\bibitem[\protect\citeauthoryear{{Cen} \& {Ostriker}}{{Cen} \&
  {Ostriker}}{1992}]{CenOstriker1992}
{Cen} R.,  {Ostriker} J.,  1992, \mn@doi [\apj] {10.1086/171482}, \href
  {https://ui.adsabs.harvard.edu/abs/1992ApJ...393...22C} {393, 22}

\bibitem[\protect\citeauthoryear{{Chy{\.z}y}}{{Chy{\.z}y}}{2008}]{chyzy2008magnetic}
{Chy{\.z}y} K.~T.,  2008, \mn@doi [\aap] {10.1051/0004-6361:20078688}, \href
  {https://ui.adsabs.harvard.edu/abs/2008A&A...482..755C} {482, 755}

\bibitem[\protect\citeauthoryear{{Crain} \& {van de Voort}}{{Crain} \& {van de
  Voort}}{2023}]{Crain2023groups}
{Crain} R.~A.,  {van de Voort} F.,  2023, \mn@doi [\araa]
  {10.1146/annurev-astro-041923-043618}, \href
  {https://ui.adsabs.harvard.edu/abs/2023ARA&A..61..473C} {61, 473}

\bibitem[\protect\citeauthoryear{{Dolag} \& {Stasyszyn}}{{Dolag} \&
  {Stasyszyn}}{2009}]{dolag2009mhd}
{Dolag} K.,  {Stasyszyn} F.,  2009, \mn@doi [\mnras]
  {10.1111/j.1365-2966.2009.15181.x}, \href
  {https://ui.adsabs.harvard.edu/abs/2009MNRAS.398.1678D} {398, 1678}

\bibitem[\protect\citeauthoryear{{Donnert}}{{Donnert}}{2014}]{donnert2014}
{Donnert} J.~M.~F.,  2014, \mn@doi [\mnras] {10.1093/mnras/stt2291}, \href
  {https://ui.adsabs.harvard.edu/abs/2014MNRAS.438.1971D} {438, 1971}

\bibitem[\protect\citeauthoryear{{Draine}}{{Draine}}{2011}]{Draine2011bookISM}
{Draine} B.~T.,  2011, {Physics of the Interstellar and Intergalactic Medium}

\bibitem[\protect\citeauthoryear{{Dubois} \& {Teyssier}}{{Dubois} \&
  {Teyssier}}{2010}]{Dubois2010}
{Dubois} Y.,  {Teyssier} R.,  2010, \mn@doi [\aap]
  {10.1051/0004-6361/200913014}, \href
  {https://ui.adsabs.harvard.edu/abs/2010A&A...523A..72D} {523, A72}

\bibitem[\protect\citeauthoryear{{Ebagezio}, {Seifried}, {Walch},
  {N{\"u}rnberger}, {Rathjen}  \& {Naab}}{{Ebagezio}
  et~al.}{2023}]{Ebagezio2023MNRAS}
{Ebagezio} S.,  {Seifried} D.,  {Walch} S.,  {N{\"u}rnberger} P.~C.,  {Rathjen}
  T.~E.,   {Naab} T.,  2023, \mn@doi [\mnras] {10.1093/mnras/stad2630}, \href
  {https://ui.adsabs.harvard.edu/abs/2023MNRAS.525.5631E} {525, 5631}

\bibitem[\protect\citeauthoryear{{Elia} et~al.,}{{Elia}
  et~al.}{2022}]{Elia2022}
{Elia} D.,  et~al., 2022, \mn@doi [\apj] {10.3847/1538-4357/aca27d}, \href
  {https://ui.adsabs.harvard.edu/abs/2022ApJ...941..162E} {941, 162}

\bibitem[\protect\citeauthoryear{{Elmegreen} \& {Scalo}}{{Elmegreen} \&
  {Scalo}}{2004}]{Elmegreen2004}
{Elmegreen} B.~G.,  {Scalo} J.,  2004, \mn@doi [\araa]
  {10.1146/annurev.astro.41.011802.094859}, \href
  {https://ui.adsabs.harvard.edu/abs/2004ARA&A..42..211E} {42, 211}

\bibitem[\protect\citeauthoryear{{Federrath} \& {Klessen}}{{Federrath} \&
  {Klessen}}{2012}]{Federrath2012}
{Federrath} C.,  {Klessen} R.~S.,  2012, \mn@doi [\apj]
  {10.1088/0004-637X/761/2/156}, \href
  {https://ui.adsabs.harvard.edu/abs/2012ApJ...761..156F} {761, 156}

\bibitem[\protect\citeauthoryear{{Gatto} et~al.,}{{Gatto}
  et~al.}{2015}]{Gatto2015}
{Gatto} A.,  et~al., 2015, \mn@doi [\mnras] {10.1093/mnras/stv324}, \href
  {https://ui.adsabs.harvard.edu/abs/2015MNRAS.449.1057G} {449, 1057}

\bibitem[\protect\citeauthoryear{{Giammaria}, {Spagna}, {Lattanzi}, {Murante},
  {Re Fiorentin}  \& {Valentini}}{{Giammaria}
  et~al.}{2021}]{Giammaria2021disks}
{Giammaria} M.,  {Spagna} A.,  {Lattanzi} M.~G.,  {Murante} G.,  {Re Fiorentin}
  P.,   {Valentini} M.,  2021, \mn@doi [\mnras] {10.1093/mnras/stab136}, \href
  {https://ui.adsabs.harvard.edu/abs/2021MNRAS.502.2251G} {502, 2251}

\bibitem[\protect\citeauthoryear{{Girma} \& {Teyssier}}{{Girma} \&
  {Teyssier}}{2024}]{Girma-Teyssier2024}
{Girma} E.,  {Teyssier} R.,  2024, \mn@doi [\mnras] {10.1093/mnras/stad3640},
  \href {https://ui.adsabs.harvard.edu/abs/2024MNRAS.527.6779G} {527, 6779}

\bibitem[\protect\citeauthoryear{{Governato}, {Mayer}  \& {Brook}}{{Governato}
  et~al.}{2008}]{2008Governato}
{Governato} F.,  {Mayer} L.,   {Brook} C.,  2008, in {Funes} J.~G.,  {Corsini}
  E.~M.,  eds,  Astronomical Society of the Pacific Conference Series Vol. 396,
  Formation and Evolution of Galaxy Disks. p.~453 (\mn@eprint {arXiv}
  {0801.1707})

\bibitem[\protect\citeauthoryear{{Grand} et~al.,}{{Grand}
  et~al.}{2017}]{Grand2017}
{Grand} R. J.~J.,  et~al., 2017, \mn@doi [\mnras] {10.1093/mnras/stx071}, \href
  {https://ui.adsabs.harvard.edu/abs/2017MNRAS.467..179G} {467, 179}

\bibitem[\protect\citeauthoryear{{Guedes}, {Callegari}, {Madau}  \&
  {Mayer}}{{Guedes} et~al.}{2011}]{guedes2011forming}
{Guedes} J.,  {Callegari} S.,  {Madau} P.,   {Mayer} L.,  2011, \mn@doi [\apj]
  {10.1088/0004-637X/742/2/76}, \href
  {https://ui.adsabs.harvard.edu/abs/2011ApJ...742...76G} {742, 76}

\bibitem[\protect\citeauthoryear{{Haid}, {Walch}, {Naab}, {Seifried}, {Mackey}
  \& {Gatto}}{{Haid} et~al.}{2016}]{Haid2016}
{Haid} S.,  {Walch} S.,  {Naab} T.,  {Seifried} D.,  {Mackey} J.,   {Gatto} A.,
   2016, \mn@doi [\mnras] {10.1093/mnras/stw1082}, \href
  {https://ui.adsabs.harvard.edu/abs/2016MNRAS.460.2962H} {460, 2962}

\bibitem[\protect\citeauthoryear{{Heesen} et~al.,}{{Heesen}
  et~al.}{2023}]{Heesen2023}
{Heesen} V.,  et~al., 2023, \mn@doi [\aap] {10.1051/0004-6361/202243328}, \href
  {https://ui.adsabs.harvard.edu/abs/2023A&A...669A...8H} {669, A8}

\bibitem[\protect\citeauthoryear{{Hernquist}}{{Hernquist}}{1993}]{hernquist1993n}
{Hernquist} L.,  1993, \mn@doi [\apjs] {10.1086/191784}, \href
  {https://ui.adsabs.harvard.edu/abs/1993ApJS...86..389H} {86, 389}

\bibitem[\protect\citeauthoryear{{Hirashima}, {Moriwaki}, {Fujii}, {Hirai},
  {Saitoh}  \& {Makino}}{{Hirashima} et~al.}{2023}]{Hirashima2023}
{Hirashima} K.,  {Moriwaki} K.,  {Fujii} M.~S.,  {Hirai} Y.,  {Saitoh} T.~R.,
  {Makino} J.,  2023, \mn@doi [\mnras] {10.1093/mnras/stad2864}, \href
  {https://ui.adsabs.harvard.edu/abs/2023MNRAS.526.4054H} {526, 4054}

\bibitem[\protect\citeauthoryear{{Hirschmann}, {Dolag}, {Saro}, {Bachmann},
  {Borgani}  \& {Burkert}}{{Hirschmann}
  et~al.}{2014}]{hirschmann2014cosmological}
{Hirschmann} M.,  {Dolag} K.,  {Saro} A.,  {Bachmann} L.,  {Borgani} S.,
  {Burkert} A.,  2014, \mn@doi [\mnras] {10.1093/mnras/stu1023}, \href
  {https://ui.adsabs.harvard.edu/abs/2014MNRAS.442.2304H} {442, 2304}

\bibitem[\protect\citeauthoryear{{Hopkins}}{{Hopkins}}{2015}]{hopkins2015new}
{Hopkins} P.~F.,  2015, \mn@doi [\mnras] {10.1093/mnras/stv195}, \href
  {https://ui.adsabs.harvard.edu/abs/2015MNRAS.450...53H} {450, 53}

\bibitem[\protect\citeauthoryear{{Hopkins}, {Quataert}  \& {Murray}}{{Hopkins}
  et~al.}{2012}]{Hopkins2012}
{Hopkins} P.~F.,  {Quataert} E.,   {Murray} N.,  2012, \mn@doi [\mnras]
  {10.1111/j.1365-2966.2012.20578.x}, \href
  {https://ui.adsabs.harvard.edu/abs/2012MNRAS.421.3488H} {421, 3488}

\bibitem[\protect\citeauthoryear{{Hopkins}, {Kere{\v{s}}}, {O{\~n}orbe},
  {Faucher-Gigu{\`e}re}, {Quataert}, {Murray}  \& {Bullock}}{{Hopkins}
  et~al.}{2014}]{hopkins2014galaxies}
{Hopkins} P.~F.,  {Kere{\v{s}}} D.,  {O{\~n}orbe} J.,  {Faucher-Gigu{\`e}re}
  C.-A.,  {Quataert} E.,  {Murray} N.,   {Bullock} J.~S.,  2014, \mn@doi
  [\mnras] {10.1093/mnras/stu1738}, \href
  {https://ui.adsabs.harvard.edu/abs/2014MNRAS.445..581H} {445, 581}

\bibitem[\protect\citeauthoryear{{Hopkins} et~al.,}{{Hopkins}
  et~al.}{2018}]{Hopkins2018fire}
{Hopkins} P.~F.,  et~al., 2018, \mn@doi [\mnras] {10.1093/mnras/sty1690}, \href
  {https://ui.adsabs.harvard.edu/abs/2018MNRAS.480..800H} {480, 800}

\bibitem[\protect\citeauthoryear{{Hopkins} et~al.,}{{Hopkins}
  et~al.}{2020}]{Hopkins2020zoom}
{Hopkins} P.~F.,  et~al., 2020, \mn@doi [\mnras] {10.1093/mnras/stz3321}, \href
  {https://ui.adsabs.harvard.edu/abs/2020MNRAS.492.3465H} {492, 3465}

\bibitem[\protect\citeauthoryear{{Hopkins}, {Butsky}, {Panopoulou}, {Ji},
  {Quataert}, {Faucher-Gigu{\`e}re}  \& {Kere{\v{s}}}}{{Hopkins}
  et~al.}{2022}]{Hopkins2022}
{Hopkins} P.~F.,  {Butsky} I.~S.,  {Panopoulou} G.~V.,  {Ji} S.,  {Quataert}
  E.,  {Faucher-Gigu{\`e}re} C.-A.,   {Kere{\v{s}}} D.,  2022, \mn@doi [\mnras]
  {10.1093/mnras/stac1791}, \href
  {https://ui.adsabs.harvard.edu/abs/2022MNRAS.tmp.1768H} {}

\bibitem[\protect\citeauthoryear{{Hopkins} et~al.,}{{Hopkins}
  et~al.}{2023}]{Hopkins2023}
{Hopkins} P.~F.,  et~al., 2023, \mn@doi [\mnras] {10.1093/mnras/stac3489},
  \href {https://ui.adsabs.harvard.edu/abs/2023MNRAS.519.3154H} {519, 3154}

\bibitem[\protect\citeauthoryear{{Hu}, {Naab}, {Glover}, {Walch}  \&
  {Clark}}{{Hu} et~al.}{2017}]{Hu2017}
{Hu} C.-Y.,  {Naab} T.,  {Glover} S. C.~O.,  {Walch} S.,   {Clark} P.~C.,
  2017, \mn@doi [\mnras] {10.1093/mnras/stx1773}, \href
  {https://ui.adsabs.harvard.edu/abs/2017MNRAS.471.2151H} {471, 2151}

\bibitem[\protect\citeauthoryear{{Katz} \& {Gunn}}{{Katz} \&
  {Gunn}}{1991}]{KatzGunn1991}
{Katz} N.,  {Gunn} J.~E.,  1991, \mn@doi [\apj] {10.1086/170367}, \href
  {https://ui.adsabs.harvard.edu/abs/1991ApJ...377..365K} {377, 365}

\bibitem[\protect\citeauthoryear{{Katz}, {Weinberg}  \& {Hernquist}}{{Katz}
  et~al.}{1996}]{katz1996cosmological}
{Katz} N.,  {Weinberg} D.~H.,   {Hernquist} L.,  1996, \mn@doi [\apj]
  {10.1086/192305}, \href
  {https://ui.adsabs.harvard.edu/abs/1996ApJS..105...19K} {105, 19}

\bibitem[\protect\citeauthoryear{{Kauffmann}, {Guiderdoni}  \&
  {White}}{{Kauffmann} et~al.}{1994}]{Kauffmann1994}
{Kauffmann} G.,  {Guiderdoni} B.,   {White} S.~D.~M.,  1994, \mn@doi [\mnras]
  {10.1093/mnras/267.4.981}, \href
  {https://ui.adsabs.harvard.edu/abs/1994MNRAS.267..981K} {267, 981}

\bibitem[\protect\citeauthoryear{{Kennicutt}}{{Kennicutt}}{1998}]{kennicutt1998global}
{Kennicutt} Robert~C. J.,  1998, \mn@doi [\apj] {10.1086/305588}, \href
  {https://ui.adsabs.harvard.edu/abs/1998ApJ...498..541K} {498, 541}

\bibitem[\protect\citeauthoryear{{Kim} \& {Ostriker}}{{Kim} \&
  {Ostriker}}{2015}]{KimOstriker2015}
{Kim} C.-G.,  {Ostriker} E.~C.,  2015, \mn@doi [\apj]
  {10.1088/0004-637X/802/2/99}, \href
  {https://ui.adsabs.harvard.edu/abs/2015ApJ...802...99K} {802, 99}

\bibitem[\protect\citeauthoryear{{Krause}}{{Krause}}{2015}]{Krause2015BGalaxy}
{Krause} M.,  2015, \mn@doi [Highlights of Astronomy]
  {10.1017/S1743921314011685}, \href
  {https://ui.adsabs.harvard.edu/abs/2015HiA....16..399K} {16, 399}

\bibitem[\protect\citeauthoryear{{Krause}}{{Krause}}{2019}]{Krause2019}
{Krause} M.,  2019, \mn@doi [Galaxies] {10.3390/galaxies7020054}, \href
  {https://ui.adsabs.harvard.edu/abs/2019Galax...7...54K} {7, 54}

\bibitem[\protect\citeauthoryear{{Kretschmer} \& {Teyssier}}{{Kretschmer} \&
  {Teyssier}}{2020}]{Kretschmer2020}
{Kretschmer} M.,  {Teyssier} R.,  2020, \mn@doi [\mnras]
  {10.1093/mnras/stz3495}, \href
  {https://ui.adsabs.harvard.edu/abs/2020MNRAS.492.1385K} {492, 1385}

\bibitem[\protect\citeauthoryear{{Krumholz} \& {McKee}}{{Krumholz} \&
  {McKee}}{2005}]{Krumholz2005}
{Krumholz} M.~R.,  {McKee} C.~F.,  2005, \mn@doi [\apj] {10.1086/431734}, \href
  {https://ui.adsabs.harvard.edu/abs/2005ApJ...630..250K} {630, 250}

\bibitem[\protect\citeauthoryear{{Lacey} \& {Silk}}{{Lacey} \&
  {Silk}}{1991}]{Lacey1991}
{Lacey} C.,  {Silk} J.,  1991, \mn@doi [\apj] {10.1086/170625}, \href
  {https://ui.adsabs.harvard.edu/abs/1991ApJ...381...14L} {381, 14}

\bibitem[\protect\citeauthoryear{{Lucas}, {Bonnell}  \& {Dale}}{{Lucas}
  et~al.}{2020}]{Lucas2020}
{Lucas} W.~E.,  {Bonnell} I.~A.,   {Dale} J.~E.,  2020, \mn@doi [\mnras]
  {10.1093/mnras/staa451}, \href
  {https://ui.adsabs.harvard.edu/abs/2020MNRAS.493.4700L} {493, 4700}

\bibitem[\protect\citeauthoryear{{Manna} \& {Roy}}{{Manna} \&
  {Roy}}{2023a}]{Manna2023}
{Manna} S.,  {Roy} S.,  2023a, \mn@doi [\apj] {10.3847/1538-4357/acaf64}, \href
  {https://ui.adsabs.harvard.edu/abs/2023ApJ...944...86M} {944, 86}

\bibitem[\protect\citeauthoryear{{Manna} \& {Roy}}{{Manna} \&
  {Roy}}{2023b}]{Manna2023B-gas}
{Manna} S.,  {Roy} S.,  2023b, \mn@doi [\apj] {10.3847/1538-4357/acaf64}, \href
  {https://ui.adsabs.harvard.edu/abs/2023ApJ...944...86M} {944, 86}

\bibitem[\protect\citeauthoryear{{Marinacci}, {Fraternali}, {Nipoti}, {Binney},
  {Ciotti}  \& {Londrillo}}{{Marinacci} et~al.}{2011}]{marinacci2011}
{Marinacci} F.,  {Fraternali} F.,  {Nipoti} C.,  {Binney} J.,  {Ciotti} L.,
  {Londrillo} P.,  2011, \mn@doi [\mnras] {10.1111/j.1365-2966.2011.18810.x},
  \href {https://ui.adsabs.harvard.edu/abs/2011MNRAS.415.1534M} {415, 1534}

\bibitem[\protect\citeauthoryear{{Marinacci}, {Pakmor}  \&
  {Springel}}{{Marinacci} et~al.}{2014}]{marinacci2014formation}
{Marinacci} F.,  {Pakmor} R.,   {Springel} V.,  2014, \mn@doi [\mnras]
  {10.1093/mnras/stt2003}, \href
  {https://ui.adsabs.harvard.edu/abs/2014MNRAS.437.1750M} {437, 1750}

\bibitem[\protect\citeauthoryear{{Marinacci} et~al.,}{{Marinacci}
  et~al.}{2018}]{Marinacci2018}
{Marinacci} F.,  et~al., 2018, \mn@doi [\mnras] {10.1093/mnras/sty2206}, \href
  {https://ui.adsabs.harvard.edu/abs/2018MNRAS.480.5113M} {480, 5113}

\bibitem[\protect\citeauthoryear{{Martin-Alvarez}, {Devriendt}, {Slyz}  \&
  {Teyssier}}{{Martin-Alvarez} et~al.}{2018}]{Martin-Alvarez2018}
{Martin-Alvarez} S.,  {Devriendt} J.,  {Slyz} A.,   {Teyssier} R.,  2018,
  \mn@doi [\mnras] {10.1093/mnras/sty1623}, \href
  {https://ui.adsabs.harvard.edu/abs/2018MNRAS.479.3343M} {479, 3343}

\bibitem[\protect\citeauthoryear{{Martin-Alvarez}, {Katz}, {Sijacki},
  {Devriendt}  \& {Slyz}}{{Martin-Alvarez} et~al.}{2021}]{Martin-Alvarez2021}
{Martin-Alvarez} S.,  {Katz} H.,  {Sijacki} D.,  {Devriendt} J.,   {Slyz} A.,
  2021, \mn@doi [\mnras] {10.1093/mnras/stab968}, \href
  {https://ui.adsabs.harvard.edu/abs/2021MNRAS.504.2517M} {504, 2517}

\bibitem[\protect\citeauthoryear{{Martizzi}, {Faucher-Gigu{\`e}re}  \&
  {Quataert}}{{Martizzi} et~al.}{2015}]{Martizzi2015}
{Martizzi} D.,  {Faucher-Gigu{\`e}re} C.-A.,   {Quataert} E.,  2015, \mn@doi
  [\mnras] {10.1093/mnras/stv562}, \href
  {https://ui.adsabs.harvard.edu/abs/2015MNRAS.450..504M} {450, 504}

\bibitem[\protect\citeauthoryear{{Mastropietro} \& {Burkert}}{{Mastropietro} \&
  {Burkert}}{2008}]{mastropietro2008simulating}
{Mastropietro} C.,  {Burkert} A.,  2008, \mn@doi [\mnras]
  {10.1111/j.1365-2966.2008.13626.x}, \href
  {https://ui.adsabs.harvard.edu/abs/2008MNRAS.389..967M} {389, 967}

\bibitem[\protect\citeauthoryear{{McKee} \& {Ostriker}}{{McKee} \&
  {Ostriker}}{1977}]{mckee1977theory}
{McKee} C.~F.,  {Ostriker} J.~P.,  1977, \mn@doi [\apj] {10.1086/155667}, \href
  {https://ui.adsabs.harvard.edu/abs/1977ApJ...218..148M} {218, 148}

\bibitem[\protect\citeauthoryear{{Mihos} \& {Hernquist}}{{Mihos} \&
  {Hernquist}}{1996}]{MihosHernquist1996}
{Mihos} J.~C.,  {Hernquist} L.,  1996, \mn@doi [\apj] {10.1086/177353}, \href
  {https://ui.adsabs.harvard.edu/abs/1996ApJ...464..641M} {464, 641}

\bibitem[\protect\citeauthoryear{{Moster}, {Macci{\`o}}, {Somerville},
  {Johansson}  \& {Naab}}{{Moster} et~al.}{2010}]{moster2010can}
{Moster} B.~P.,  {Macci{\`o}} A.~V.,  {Somerville} R.~S.,  {Johansson} P.~H.,
  {Naab} T.,  2010, \mn@doi [\mnras] {10.1111/j.1365-2966.2009.16190.x}, \href
  {https://ui.adsabs.harvard.edu/abs/2010MNRAS.403.1009M} {403, 1009}

\bibitem[\protect\citeauthoryear{{Murante}, {Monaco}, {Giovalli}, {Borgani}  \&
  {Diaferio}}{{Murante} et~al.}{2010}]{murante2010subresolution}
{Murante} G.,  {Monaco} P.,  {Giovalli} M.,  {Borgani} S.,   {Diaferio} A.,
  2010, \mn@doi [\mnras] {10.1111/j.1365-2966.2010.16567.x}, \href
  {https://ui.adsabs.harvard.edu/abs/2010MNRAS.405.1491M} {405, 1491}

\bibitem[\protect\citeauthoryear{{Murante}, {Monaco}, {Borgani}, {Tornatore},
  {Dolag}  \& {Goz}}{{Murante} et~al.}{2015}]{murante2015simulating}
{Murante} G.,  {Monaco} P.,  {Borgani} S.,  {Tornatore} L.,  {Dolag} K.,
  {Goz} D.,  2015, \mn@doi [\mnras] {10.1093/mnras/stu2400}, \href
  {https://ui.adsabs.harvard.edu/abs/2015MNRAS.447..178M} {447, 178}

\bibitem[\protect\citeauthoryear{{Naab} \& {Ostriker}}{{Naab} \&
  {Ostriker}}{2017}]{Naab2017}
{Naab} T.,  {Ostriker} J.~P.,  2017, \mn@doi [\araa]
  {10.1146/annurev-astro-081913-040019}, \href
  {https://ui.adsabs.harvard.edu/abs/2017ARA&A..55...59N} {55, 59}

\bibitem[\protect\citeauthoryear{{Navarro} \& {White}}{{Navarro} \&
  {White}}{1993}]{NavarroWhite1993}
{Navarro} J.~F.,  {White} S.~D.~M.,  1993, \mn@doi [\mnras]
  {10.1093/mnras/265.2.271}, \href
  {https://ui.adsabs.harvard.edu/abs/1993MNRAS.265..271N} {265, 271}

\bibitem[\protect\citeauthoryear{{Navarro} \& {White}}{{Navarro} \&
  {White}}{1994}]{NavarroWhite1994}
{Navarro} J.~F.,  {White} S. D.~M.,  1994, \mn@doi [\mnras]
  {10.1093/mnras/267.2.401}, \href
  {https://ui.adsabs.harvard.edu/abs/1994MNRAS.267..401N} {267, 401}

\bibitem[\protect\citeauthoryear{{Niklas} \& {Beck}}{{Niklas} \&
  {Beck}}{1997}]{Niklas1997}
{Niklas} S.,  {Beck} R.,  1997, \aap, \href
  {https://ui.adsabs.harvard.edu/abs/1997A&A...320...54N} {320, 54}

\bibitem[\protect\citeauthoryear{{Ntormousi}, {Tassis}, {Del Sordo},
  {Fragkoudi}  \& {Pakmor}}{{Ntormousi} et~al.}{2020}]{Ntormousi2020}
{Ntormousi} E.,  {Tassis} K.,  {Del Sordo} F.,  {Fragkoudi} F.,   {Pakmor} R.,
  2020, \mn@doi [\aap] {10.1051/0004-6361/202037835}, \href
  {https://ui.adsabs.harvard.edu/abs/2020A&A...641A.165N} {641, A165}

\bibitem[\protect\citeauthoryear{{Oser}, {Ostriker}, {Naab}, {Johansson}  \&
  {Burkert}}{{Oser} et~al.}{2010}]{oser2010two}
{Oser} L.,  {Ostriker} J.~P.,  {Naab} T.,  {Johansson} P.~H.,   {Burkert} A.,
  2010, \mn@doi [\apj] {10.1088/0004-637X/725/2/2312}, \href
  {https://ui.adsabs.harvard.edu/abs/2010ApJ...725.2312O} {725, 2312}

\bibitem[\protect\citeauthoryear{{Pakmor} \& {Springel}}{{Pakmor} \&
  {Springel}}{2013}]{pakmor2013simulations}
{Pakmor} R.,  {Springel} V.,  2013, \mn@doi [\mnras] {10.1093/mnras/stt428},
  \href {https://ui.adsabs.harvard.edu/abs/2013MNRAS.432..176P} {432, 176}

\bibitem[\protect\citeauthoryear{{Pakmor} et~al.,}{{Pakmor}
  et~al.}{2017}]{Pakmor2017}
{Pakmor} R.,  et~al., 2017, \mn@doi [\mnras] {10.1093/mnras/stx1074}, \href
  {https://ui.adsabs.harvard.edu/abs/2017MNRAS.469.3185P} {469, 3185}

\bibitem[\protect\citeauthoryear{{Pakmor} et~al.,}{{Pakmor}
  et~al.}{2024}]{Pakmor2024}
{Pakmor} R.,  et~al., 2024, \mn@doi [\mnras] {10.1093/mnras/stae112}, \href
  {https://ui.adsabs.harvard.edu/abs/2024MNRAS.528.2308P} {528, 2308}

\bibitem[\protect\citeauthoryear{{Pillepich} et~al.,}{{Pillepich}
  et~al.}{2018}]{Pillepich2018}
{Pillepich} A.,  et~al., 2018, \mn@doi [\mnras] {10.1093/mnras/stx2656}, \href
  {https://ui.adsabs.harvard.edu/abs/2018MNRAS.473.4077P} {473, 4077}

\bibitem[\protect\citeauthoryear{{Ramesh}, {Nelson}, {Heesen}  \&
  {Br{\"u}ggen}}{{Ramesh} et~al.}{2023}]{Ramesh2023MNRAS}
{Ramesh} R.,  {Nelson} D.,  {Heesen} V.,   {Br{\"u}ggen} M.,  2023, \mn@doi
  [\mnras] {10.1093/mnras/stad3104}, \href
  {https://ui.adsabs.harvard.edu/abs/2023MNRAS.526.5483R} {526, 5483}

\bibitem[\protect\citeauthoryear{{Rieder} \& {Teyssier}}{{Rieder} \&
  {Teyssier}}{2016}]{Rieder2016}
{Rieder} M.,  {Teyssier} R.,  2016, \mn@doi [\mnras] {10.1093/mnras/stv2985},
  \href {https://ui.adsabs.harvard.edu/abs/2016MNRAS.457.1722R} {457, 1722}

\bibitem[\protect\citeauthoryear{{Rieder} \& {Teyssier}}{{Rieder} \&
  {Teyssier}}{2017a}]{Rieder2017a}
{Rieder} M.,  {Teyssier} R.,  2017a, \mn@doi [\mnras] {10.1093/mnras/stx1670},
  \href {https://ui.adsabs.harvard.edu/abs/2017MNRAS.471.2674R} {471, 2674}

\bibitem[\protect\citeauthoryear{{Rieder} \& {Teyssier}}{{Rieder} \&
  {Teyssier}}{2017b}]{Rieder2017b}
{Rieder} M.,  {Teyssier} R.,  2017b, \mn@doi [\mnras] {10.1093/mnras/stx2276},
  \href {https://ui.adsabs.harvard.edu/abs/2017MNRAS.472.4368R} {472, 4368}

\bibitem[\protect\citeauthoryear{{Robitaille} \& {Whitney}}{{Robitaille} \&
  {Whitney}}{2010}]{robitaille2010present}
{Robitaille} T.~P.,  {Whitney} B.~A.,  2010, \mn@doi [\apjl]
  {10.1088/2041-8205/710/1/L11}, \href
  {https://ui.adsabs.harvard.edu/abs/2010ApJ...710L..11R} {710, L11}

\bibitem[\protect\citeauthoryear{{Ruzmaikin}, {Sokolov}  \&
  {Shukurov}}{{Ruzmaikin} et~al.}{1988}]{ruzmaikin1988magnetism}
{Ruzmaikin} A.,  {Sokolov} D.,   {Shukurov} A.,  1988, \mn@doi [\nat]
  {10.1038/336341a0}, \href
  {https://ui.adsabs.harvard.edu/abs/1988Natur.336..341R} {336, 341}

\bibitem[\protect\citeauthoryear{{Salpeter}}{{Salpeter}}{1955}]{salpeter1955luminosity}
{Salpeter} E.~E.,  1955, \mn@doi [\apj] {10.1086/145971}, \href
  {https://ui.adsabs.harvard.edu/abs/1955ApJ...121..161S} {121, 161}

\bibitem[\protect\citeauthoryear{{Scalo} \& {Elmegreen}}{{Scalo} \&
  {Elmegreen}}{2004}]{Scalo2004}
{Scalo} J.,  {Elmegreen} B.~G.,  2004, \mn@doi [\araa]
  {10.1146/annurev.astro.42.120403.143327}, \href
  {https://ui.adsabs.harvard.edu/abs/2004ARA&A..42..275S} {42, 275}

\bibitem[\protect\citeauthoryear{{Schaye} et~al.,}{{Schaye}
  et~al.}{2015}]{Schaye2015}
{Schaye} J.,  et~al., 2015, \mn@doi [\mnras] {10.1093/mnras/stu2058}, \href
  {https://ui.adsabs.harvard.edu/abs/2015MNRAS.446..521S} {446, 521}

\bibitem[\protect\citeauthoryear{{Schleicher} \& {Beck}}{{Schleicher} \&
  {Beck}}{2013}]{Schleicher2013}
{Schleicher} D. R.~G.,  {Beck} R.,  2013, \mn@doi [\aap]
  {10.1051/0004-6361/201321707}, \href
  {https://ui.adsabs.harvard.edu/abs/2013A&A...556A.142S} {556, A142}

\bibitem[\protect\citeauthoryear{{Schmidt}}{{Schmidt}}{1959}]{schmidt1959rate}
{Schmidt} M.,  1959, \mn@doi [\apj] {10.1086/146614}, \href
  {https://ui.adsabs.harvard.edu/abs/1959ApJ...129..243S} {129, 243}

\bibitem[\protect\citeauthoryear{{Semenov}, {Kravtsov}  \& {Gnedin}}{{Semenov}
  et~al.}{2018}]{Semenov2018}
{Semenov} V.~A.,  {Kravtsov} A.~V.,   {Gnedin} N.~Y.,  2018, \mn@doi [\apj]
  {10.3847/1538-4357/aac6eb}, \href
  {https://ui.adsabs.harvard.edu/abs/2018ApJ...861....4S} {861, 4}

\bibitem[\protect\citeauthoryear{{Somerville} \& {Dav{\'e}}}{{Somerville} \&
  {Dav{\'e}}}{2015}]{Somerville2015}
{Somerville} R.~S.,  {Dav{\'e}} R.,  2015, \mn@doi [\araa]
  {10.1146/annurev-astro-082812-140951}, \href
  {https://ui.adsabs.harvard.edu/abs/2015ARA&A..53...51S} {53, 51}

\bibitem[\protect\citeauthoryear{{Springel}}{{Springel}}{2005}]{springel2005gadget}
{Springel} V.,  2005, \mn@doi [\mnras] {10.1111/j.1365-2966.2005.09655.x},
  \href {https://ui.adsabs.harvard.edu/abs/2005MNRAS.364.1105S} {364, 1105}

\bibitem[\protect\citeauthoryear{{Springel} \& {Hernquist}}{{Springel} \&
  {Hernquist}}{2003}]{springel2003cosmological}
{Springel} V.,  {Hernquist} L.,  2003, \mn@doi [\mnras]
  {10.1046/j.1365-8711.2003.06206.x}, \href
  {https://ui.adsabs.harvard.edu/abs/2003MNRAS.339..289S} {339, 289}

\bibitem[\protect\citeauthoryear{{Springel}, {Yoshida}  \& {White}}{{Springel}
  et~al.}{2001}]{springel2001gadget}
{Springel} V.,  {Yoshida} N.,   {White} S. D.~M.,  2001, \mn@doi [\na]
  {10.1016/S1384-1076(01)00042-2}, \href
  {https://ui.adsabs.harvard.edu/abs/2001NewA....6...79S} {6, 79}

\bibitem[\protect\citeauthoryear{{Steinmetz} \& {Muller}}{{Steinmetz} \&
  {Muller}}{1995}]{Steinmetz1995}
{Steinmetz} M.,  {Muller} E.,  1995, \mn@doi [\mnras]
  {10.1093/mnras/276.2.549}, \href
  {https://ui.adsabs.harvard.edu/abs/1995MNRAS.276..549S} {276, 549}

\bibitem[\protect\citeauthoryear{{Steinwandel}, {Beck}, {Arth}, {Dolag},
  {Moster}  \& {Nielaba}}{{Steinwandel} et~al.}{2019}]{Ulli}
{Steinwandel} U.~P.,  {Beck} M.~C.,  {Arth} A.,  {Dolag} K.,  {Moster} B.~P.,
  {Nielaba} P.,  2019, \mn@doi [\mnras] {10.1093/mnras/sty3083}, \href
  {https://ui.adsabs.harvard.edu/abs/2019MNRAS.483.1008S} {483, 1008}

\bibitem[\protect\citeauthoryear{{Steinwandel}, {Dolag}, {Lesch}, {Moster},
  {Burkert}  \& {Prieto}}{{Steinwandel} et~al.}{2020}]{Ulli-Bwinds}
{Steinwandel} U.~P.,  {Dolag} K.,  {Lesch} H.,  {Moster} B.~P.,  {Burkert} A.,
   {Prieto} A.,  2020, \mn@doi [\mnras] {10.1093/mnras/staa817}, \href
  {https://ui.adsabs.harvard.edu/abs/2020MNRAS.tmp..740S} {}

\bibitem[\protect\citeauthoryear{{Su}, {Hopkins}, {Hayward},
  {Faucher-Gigu{\`e}re}, {Kere{\v{s}}}, {Ma}  \& {Robles}}{{Su}
  et~al.}{2017}]{Su2017}
{Su} K.-Y.,  {Hopkins} P.~F.,  {Hayward} C.~C.,  {Faucher-Gigu{\`e}re} C.-A.,
  {Kere{\v{s}}} D.,  {Ma} X.,   {Robles} V.~H.,  2017, \mn@doi [\mnras]
  {10.1093/mnras/stx1463}, \href
  {https://ui.adsabs.harvard.edu/abs/2017MNRAS.471..144S} {471, 144}

\bibitem[\protect\citeauthoryear{{Tabatabaei}, {Krause}, {Fletcher}  \&
  {Beck}}{{Tabatabaei} et~al.}{2008}]{Tabatabaei2008}
{Tabatabaei} F.~S.,  {Krause} M.,  {Fletcher} A.,   {Beck} R.,  2008, \mn@doi
  [\aap] {10.1051/0004-6361:200810590}, \href
  {https://ui.adsabs.harvard.edu/abs/2008A&A...490.1005T} {490, 1005}

\bibitem[\protect\citeauthoryear{{Tabatabaei} et~al.,}{{Tabatabaei}
  et~al.}{2017}]{2017ApJ...836..185T}
{Tabatabaei} F.~S.,  et~al., 2017, \mn@doi [\apj]
  {10.3847/1538-4357/836/2/185}, \href
  {https://ui.adsabs.harvard.edu/abs/2017ApJ...836..185T} {836, 185}

\bibitem[\protect\citeauthoryear{{Tabatabaei}, {Minguez}, {Prieto}  \&
  {Fern{\'a}ndez-Ontiveros}}{{Tabatabaei} et~al.}{2018}]{2018NatAs...2...83T}
{Tabatabaei} F.~S.,  {Minguez} P.,  {Prieto} M.~A.,   {Fern{\'a}ndez-Ontiveros}
  J.~A.,  2018, \mn@doi [Nature Astronomy] {10.1038/s41550-017-0298-7}, \href
  {https://ui.adsabs.harvard.edu/abs/2018NatAs...2...83T} {2, 83}

\bibitem[\protect\citeauthoryear{{Valentini}, {Murante}, {Borgani}, {Monaco},
  {Bressan}  \& {Beck}}{{Valentini} et~al.}{2017}]{valentini2017effect}
{Valentini} M.,  {Murante} G.,  {Borgani} S.,  {Monaco} P.,  {Bressan} A.,
  {Beck} A.~M.,  2017, \mn@doi [\mnras] {10.1093/mnras/stx1352}, \href
  {https://ui.adsabs.harvard.edu/abs/2017MNRAS.470.3167V} {470, 3167}

\bibitem[\protect\citeauthoryear{{Valentini}, {Bressan}, {Borgani}, {Murante},
  {Girardi}  \& {Tornatore}}{{Valentini} et~al.}{2018}]{Valentini2018}
{Valentini} M.,  {Bressan} A.,  {Borgani} S.,  {Murante} G.,  {Girardi} L.,
  {Tornatore} L.,  2018, \mn@doi [\mnras] {10.1093/mnras/sty1896}, \href
  {https://ui.adsabs.harvard.edu/abs/2018MNRAS.480..722V} {480, 722}

\bibitem[\protect\citeauthoryear{{Valentini} et~al.,}{{Valentini}
  et~al.}{2023}]{Valentini2023}
{Valentini} M.,  et~al., 2023, \mn@doi [\mnras] {10.1093/mnras/stac2110}, \href
  {https://ui.adsabs.harvard.edu/abs/2023MNRAS.518.1128V} {518, 1128}

\bibitem[\protect\citeauthoryear{{Vogelsberger} et~al.,}{{Vogelsberger}
  et~al.}{2014}]{Vogelsberger2014}
{Vogelsberger} M.,  et~al., 2014, \mn@doi [\nat] {10.1038/nature13316}, \href
  {https://ui.adsabs.harvard.edu/abs/2014Natur.509..177V} {509, 177}

\bibitem[\protect\citeauthoryear{{Walch} \& {Naab}}{{Walch} \&
  {Naab}}{2015}]{WalchNaab2015}
{Walch} S.,  {Naab} T.,  2015, \mn@doi [\mnras] {10.1093/mnras/stv1155}, \href
  {https://ui.adsabs.harvard.edu/abs/2015MNRAS.451.2757W} {451, 2757}

\bibitem[\protect\citeauthoryear{{Wang} \& {Abel}}{{Wang} \&
  {Abel}}{2009}]{Wang2009mhd}
{Wang} P.,  {Abel} T.,  2009, \mn@doi [\apj] {10.1088/0004-637X/696/1/96},
  \href {https://ui.adsabs.harvard.edu/abs/2009ApJ...696...96W} {696, 96}

\bibitem[\protect\citeauthoryear{{Wang}, {Dutton}, {Stinson}, {Macci{\`o}},
  {Penzo}, {Kang}, {Keller}  \& {Wadsley}}{{Wang} et~al.}{2015}]{wang2015nihao}
{Wang} L.,  {Dutton} A.~A.,  {Stinson} G.~S.,  {Macci{\`o}} A.~V.,  {Penzo} C.,
   {Kang} X.,  {Keller} B.~W.,   {Wadsley} J.,  2015, \mn@doi [\mnras]
  {10.1093/mnras/stv1937}, \href
  {https://ui.adsabs.harvard.edu/abs/2015MNRAS.454...83W} {454, 83}

\bibitem[\protect\citeauthoryear{{White} \& {Rees}}{{White} \&
  {Rees}}{1978}]{WhiteRees1978}
{White} S.~D.~M.,  {Rees} M.~J.,  1978, \mn@doi [\mnras]
  {10.1093/mnras/183.3.341}, \href
  {https://ui.adsabs.harvard.edu/abs/1978MNRAS.183..341W} {183, 341}

\bibitem[\protect\citeauthoryear{{Wibking} \& {Krumholz}}{{Wibking} \&
  {Krumholz}}{2023a}]{Wibking2023BGal}
{Wibking} B.~D.,  {Krumholz} M.~R.,  2023a, \mn@doi [\mnras]
  {10.1093/mnras/stac2648}, \href
  {https://ui.adsabs.harvard.edu/abs/2023MNRAS.521.5972W} {521, 5972}

\bibitem[\protect\citeauthoryear{{Wibking} \& {Krumholz}}{{Wibking} \&
  {Krumholz}}{2023b}]{Wibking2023}
{Wibking} B.~D.,  {Krumholz} M.~R.,  2023b, \mn@doi [\mnras]
  {10.1093/mnras/stac2648}, \href
  {https://ui.adsabs.harvard.edu/abs/2023MNRAS.521.5972W} {521, 5972}

\bibitem[\protect\citeauthoryear{{Zari}, {Frankel}  \& {Rix}}{{Zari}
  et~al.}{2023}]{Zari2023}
{Zari} E.,  {Frankel} N.,   {Rix} H.-W.,  2023, \mn@doi [\aap]
  {10.1051/0004-6361/202244194}, \href
  {https://ui.adsabs.harvard.edu/abs/2023A&A...669A..10Z} {669, A10}

\makeatother
\end{thebibliography}

% if your bibtex file is called example.bib

%\begin{thebibliography}{99}
%\end{thebibliography}

%%%%%%%%%%%%%%%%%%%%%%%%%%%%%%%%%%%%%%%%%%%%%%%%%%

%%%%%%%%%%%%%%%%% APPENDICES %%%%%%%%%%%%%%%%%%%%%

%%%%%%%%%%%%%%%%%%%%%%%%%%%%%%%%%%%%%%%%%%%%%%%%%%

% Don't change these lines
\bsp	% typesetting comment
\label{lastpage}
\end{document}